\documentclass[preprint,superscriptaddress,nofootinbib,showkeys,showpacs,tightenlines]{revtex4}  
\usepackage{graphicx}
\usepackage{dcolumn}
\newcommand{\be}{\begin{equation}}
\newcommand{\ee}{\end{equation}}
\newcommand{\bee}{\begin{eqnarray}}
\newcommand{\eee}{\end{eqnarray}}
\newcommand{\eq}{\end{quote}}
\newcommand{\nn}{\nonumber}
\newcommand{\Slash}[1]{\ooalign{\hfil/\hfil\crcr$#1$}}

\def\gsim{\displaystyle\mathop{>}_{\sim}}
\def\lsim{\displaystyle\mathop{<}_{\sim}}
\begin{document}      
\preprint{PNU-NTG-6/2005}
\title{$\Lambda(1520,3/2^{-})$ photoproduction reaction via $\gamma
N\to K\Lambda(1520)$}
\author{Seung-Il Nam}
\email{sinam@rcnp.osaka-u.ac.jp}
\affiliation{Research Center for Nuclear Physics (RCNP), Osaka
  University, Ibaraki, Osaka
567-0047, Japan}
\affiliation{Department of Physics and Nuclear Physics \& Radiation Technology Institute (NuRI),
Pusan National University, Busan 609-735, Republic of Korea} 
\author{Atsushi Hosaka}
\email{hosaka@rcnp.osaka-u.ac.jp}
\affiliation{Research Center for Nuclear Physics (RCNP), Osaka
  University, Ibaraki, Osaka
567-0047, Japan}
\author{Hyun-Chul Kim}
\email{hchkim@pusan.ac.kr}
\affiliation{Department of Physics and Nuclear Physics \& Radiation Technology Institute (NuRI),
Pusan National University, Busan 609-735, Republic of Korea} 
\date{\today}
\begin{abstract}
We investigate $\Lambda(1520,3/2^{-},D_{03})$ photoproduction
via the $\gamma  N\to K\Lambda^{*}$ process. Using effective
Lagrangians, we compute the total and differential cross sections.
The dependence on the momentum transfer for the photoproduction at the
tree-level is also examined.  We find that the total cross sections
for the proton target are well reproduced as compared with the
experimental data. It turns out that the total cross sections for the
neutron target are significantly smaller than those for the proton
one. We also compare the present results with the $\gamma N\to
\bar{K}\Theta^+$ reaction in order to extract information of
$\Theta^+$.  The role of $K^*$--exchange in the production 
reaction is also discussed.    
\end{abstract}
\pacs{13.75.Cs, 14.20.-c}
\keywords{$\Lambda(1520)$, Spin-3/2, Photoproduction}
\maketitle
\section{introduction}
Recent interest in excited baryons has been largely motivated by
new experimental developments~\cite{experiment}:
The observation of the exotic $\Theta^+$ resonance of strangeness
$S= +1$ has triggered diverse activities in both experimental
and theoretical studies.  The finding of the $\Theta^+$ has renewed 
interest in baryon spectroscopy.  For instance, properties
of the $\Lambda(1405)$ has been reanalyzed, based on the idea of chiral 
perturbation theory and of dynamical generation from (anti)
kaon-nucleon scattering.  A meson-baryon bound-state picture suggests
another type of the multi-quark structure.  A spin-$3/2^-$ partner of
this resonance, i.e. $\Lambda(1520)$ $(\equiv\Lambda^*)$ whose mass 
is similar to that of $\Theta^+$ but strangeness is opposite is yet
another interesting resonance.  It can be produced simultaneously in the
$\Theta^+$ photoproduction from the deuteron target.  The LEPS
collaboration is searching for the $\Theta^+$ associated with the
production of the $\Lambda^*$ in photoproduction off the
deuteron~\cite{Nakano:chiral05}.  Since the measurement of the $\Lambda^*$
can be performed much more reliably, the detailed understanding of the
production mechanism of this resonance would be useful to extract
information of the $\Theta^+$. 

As far as the experimental data of the $\Lambda^*$ production are
concerned, there are experiments reported so far:  Boyarski
(photoproduction){\it et 
al.}~\cite{Boyarski:1970yc}, the Daresbury group
(photoproduction)~\cite{Barber:1980zv}, 
and the CLAS collaboration (electroproduction)~\cite{Barrow:2001ds}.
However, these two production mechanisms showed rather different
tendencies:  While 
in Ref.~\cite{Boyarski:1970yc} and in the Daresbury experiment
$K^*$--exchange is known to be dominant in the $t$--channel, the CLAS 
experiment indicates that pseudoscalar $K$--exchange governs the
process.  Moreover, the kinematical regions of these experiments are
different, so that a mere comparison is not meaningful.  

In the present work, we investigate the photoproduction of the $\Lambda^*$
near the threshold.  Based on the effective Lagrangian for 
meson-baryon vertices, we use the Born approximation.  We introduce
form factors at the vertices, which reflect the internal structure
of hadrons but bring in model dependence.  However, there is a caveat:
Introducing the form factors violates the gauge invariance of the
electromagnetic interaction, which causes the Ward-Takahashi identity to
be broken.  Thus, we have to take care of the form factors to
restore the gauge invariance.  Since there is no unique theoretical way to
introduce the gauge-invariant form factors, we shall adopt 
the prescriptions discussed in
Refs.~\cite{Ohta:ji,Haberzettl:1998eq,Davidson:2001qs}. 

In the present approach, we treat the $\Lambda^*$ with
spin  $J = 3/2$ in the Rarita-Schwinger
formalism~\cite{Rarita:mf,Read:ye,Johnson:1960vt,Velo:ur,
Pascalutsa:1998pw,Hoehler:gb,Nath:wp,Hagen:ea}.  Since we consider the
production of the real $\Lambda^*$, the uncertainty of the
off-shell parameter in the Rarita-Schwinger parameterization can be 
minimized.  In the present calculation, we consider both $K$-- and
$K^*$--exchanges in the $t$--channel.  They show very different
behaviors for spin-dependent quantities, which will be useful to
study the mechanism of the photoproduction.  Unknown parameters such
as strong coupling constants and magnetic moments of the $\Lambda^*$
will be adjusted so as to reproduce experimental data, being guided
by the quark model.  We shall consider the photo-reactions for both proton and 
neutron targets in order to study the role of the isospin in
understanding the reaction mechanism.

The outline of the paper is as follows:
In Section 2, we describe the effective Lagrangians for various
meson-baryon vertices and construct the invariant amplitudes. 
In Section 3, we demonstrate numerical results for total and
differential cross sections both for the proton and neutron targets.
Theoretical predictions are then compared with the existing 
experimental data.  In the final Section, we will summarize the present
work with some discussions.
\section{General formalism}
\begin{figure}[tbh]
\includegraphics{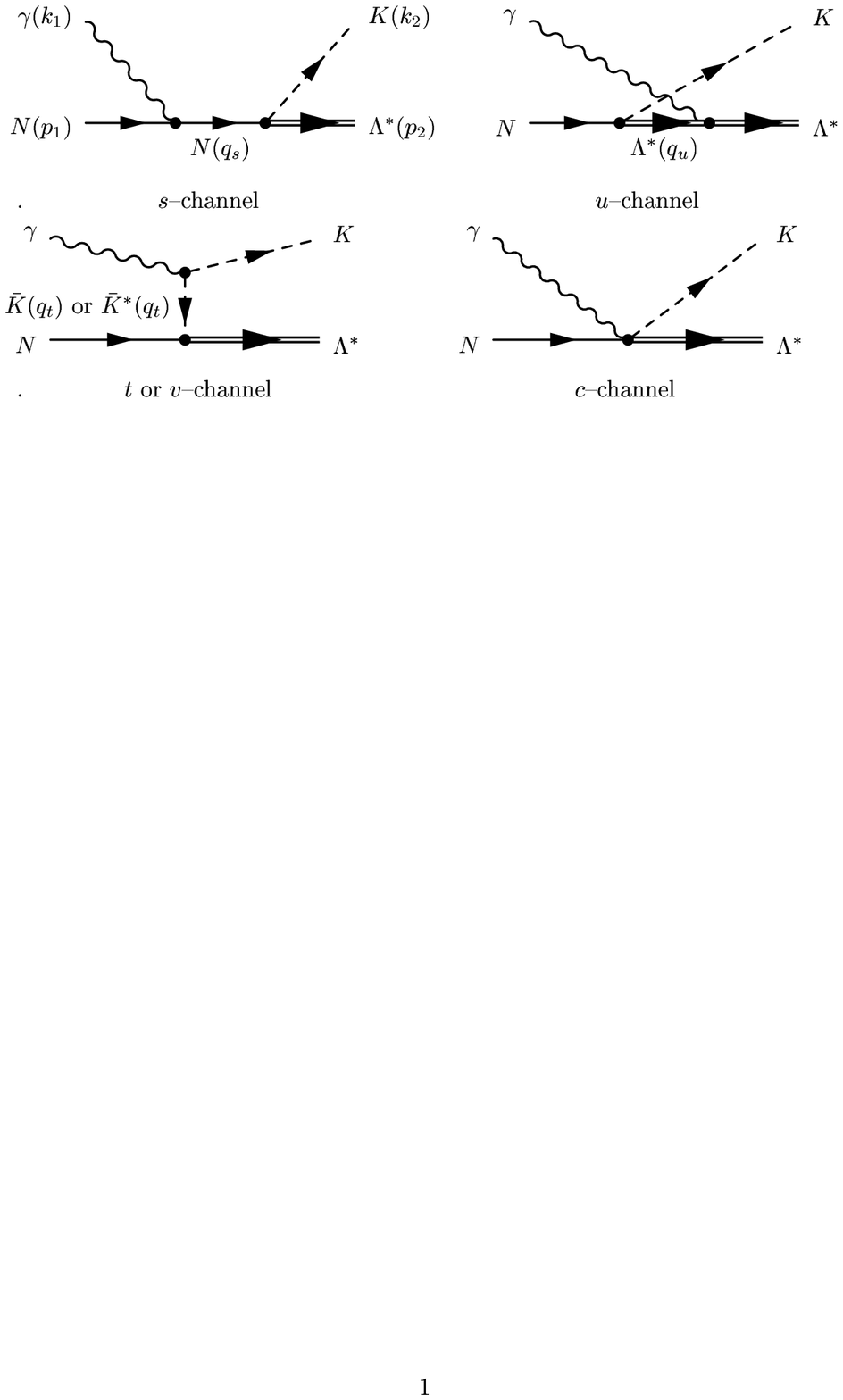}
\caption{The Feynmann diagrams}
\label{fig0}
\end{figure} 
We begin with the effective Lagrangians relevant to the $\gamma N\to
K\Lambda^{*}$ process as depicted in Fig.~\ref{fig0}.  We define the
momenta of photon, pseudo-scalar kaon $K$, vector meson $K^*$, nucleon
and $\Lambda^*$ in the figure. For convenience, vector
$K^*$--exchange in the $t$--channel and contact diagrams will be
called as the $v$--channel (vector channel) and $c$--channel
(contact-term channel), respectively.  We need to consider all diagrams
shown in Fig.~\ref{fig0} for the proton target, whereas only the magnetic
term in the $s$--channel, $K^*$--exchange in the $v$ and $u$--channels
are required for the neutron target. In order to formulate the
effective Lagrangians including spin-3/2 particles, we employ the
Rarita-Schwinger (RS) field which we summarize in the
Appendix.   

The relevant effective Lagrangians are given as :  
\bee
\mathcal{L}_{\gamma NN}
&=&-e\bar{N}\left(\gamma_{\mu}+i\frac{\kappa_{N}}{2M_{N}}
\sigma_{\mu\nu}k^{\nu}_{1}\right)  
A^{\mu}N\,+{\rm
h.c.},\nn\\
\mathcal{L}_{\gamma KK}&=&
ie\left\{ 
(\partial^{\mu}K^{\dagger})K-(\partial^{\mu}K)K^{\dagger}
\right\}A_{\mu},\nn\\
\mathcal{L}_{\gamma
\Lambda^*\Lambda^*}&=&-\bar{\Lambda}^{*\mu}
\left\{\left(-F_{1}\Slash{A}g_{\mu\nu}+F_3\Slash{A}\frac{k_{1
\mu}k_{1
\nu}}{2M^{2}_{\Lambda^*}}\right)-\frac{\Slash{k}_{1}\Slash{A}}
{2M_{\Lambda^*}}\left(-F_{2}g_{\mu\nu}+F_4\frac{k_{1\mu}k_{1 \nu}}
{2M^{2}_{\Lambda^*}}\right)\right\}\Lambda^{*\nu}\,+{\rm 
h.c.},\nn\\
\mathcal{L}_{\gamma
  KK^{*}}&=&g_{\gamma
  KK^{*}}\epsilon_{\mu\nu\sigma\rho}(\partial^{\mu}A^{\nu})
(\partial^{\sigma}K)K^{*\rho}\,+{\rm 
h.c.},\nn\\
\mathcal{L}_{KN\Lambda^*}&=&
\frac{g_{KN\Lambda^*}}{M_{K}}\bar{\Lambda}^{*\mu}
\Theta_{\mu\nu}(A,Z)(\partial^{\nu}K){\gamma}_{5}N\,+{\rm 
h.c.},\nn\\
\mathcal{L}_{K^{*}N\Lambda^*}&=&-\frac{ig_{K^{*}N\Lambda^*}}
{M_{K^*}}\bar{\Lambda}^{*\mu}\gamma^{\nu}(\partial_{\mu}
K^{*}_{\nu}-\partial_{\nu}K^{*}_{\mu})N+{\rm 
h.c.},\nn\\ \mathcal{L}_{\gamma
KN\Lambda^*}&=&-i\frac{eg_{KN\Lambda^*}}{M_{K}}\bar{\Lambda}^{*\mu}
A_{\mu}K{\gamma}_{5}N\,+{\rm h.c.},
\label{Lagrangian}
\eee
where $N$, $\Lambda^*_{\mu}$, $K$ and $A^{\mu}$ denote the nucleon,
$\Lambda^*$, pseudoscalar kaon and photon fields, respectively.  The
interaction for the $K^{*}N\Lambda^*$ vertex is taken 
from Ref.~\cite{Machleidt:1987hj}.  As for the $\gamma \Lambda^*
\Lambda^*$ vertex in the $u$--channel, we utilize the effective
interaction suggested by Ref.~\cite{gourdin} which contains four form
factors of different multipoles.  We ignore the electric coupling
$F_1$, since the $\Lambda^*$ is neutral.  We also neglect $F_3$ and
$F_4$ terms, assuming that higher multipole terms are less
important. Hence, for the photon coupling to $\Lambda^*$, we consider
only the magnetic coupling term $F_2$ whose strength is proportional
to the anomalous magnetic moment of the $\Lambda^*$,
i.e. $\kappa_{\Lambda^*}$ which is treated as a free parameter. The
off-shell term $\Theta_{\mu\nu}(A,Z)$ of the spin-3/2 particle
is defined in general as follows~\cite{Nath:wp,Hagen:ea}:  
\be
\Theta_{\mu\nu}(A,Z) = g_{\mu\nu}+\left\{\frac{1}{2}
\left(1+4Z)A+Z\right)\right\}\gamma_{\mu}\gamma_{\nu}. 
\label{theta} 
\ee
If we choose $A=-1$~\cite{Read:ye,Nath:wp,Hagen:ea}, we can
rewrite Eq.~(\ref{theta}) in the following form with a new parameter
$X=-(Z+1/2)$:
\be
\Theta_{\mu\nu}(X)=g_{\mu\nu}+X\gamma_{\mu}\gamma_{\nu},
\ee
where $X$ is regarded as a free parameter in the present work. 

In order to determine the coupling constant $g_{KN\Lambda^*}$, we make 
use of the full width $\Gamma_{\Lambda} = 15.6$ MeV and the
branching 
ratio 0.45 for the decay
$\Lambda^*\to\bar{K}N$~\cite{Eidelman:2004wy}.  The coupling constant
$KN\Lambda^*$ can be obtained by the following relation : 
\bee
g_{KN\Lambda^*}=\left\{\frac{P_{3}}{4\pi
  M^{2}_{\Lambda^{*}}M^{2}_{K}{\Gamma}_{\Lambda^{*}}}\left(\frac{1}{4}
\sum_{\rm
spin}|\mathcal{M'}|^{2}\right)\right\}^{-\frac{1}{2}},\,\,\,\,\,i\mathcal{M'}  
&=&\bar{u}(P_2)\gamma_{5}P_{3}^{\mu}u_{\mu}(P_1),
\label{coupling}
\eee
where $P_1$, $P_2$ and $P_3$ stand for the momenta of $\Lambda^{*}$, $N$
and $\bar{K}$, respectively  for the two-body decay 
$\Lambda^{*}\to\bar{K}N$ in the center of mass frame.  Thus, we obtain
$g_{KN\Lambda^*}\sim 11$. As for the $K^*N\Lambda^*$ coupling
constant, we will choose the values of $|g_{K^*N\Lambda^*}|=0$ and
$|g_{K^*N\Lambda^*}|=11$ for the numerical calculation. In the
non-relativistic quark model, if $\Lambda^*$ is described as a
$p$--wave excitation of flavor-singlet spin-3/2 state, it is shown
that the strength of the $K^*N\Lambda^*$ coupling constant is of the
same order as that of $KN\Lambda^*$ or even larger than that. The
coupling constant of the $g_{\gamma K^{*}K}$ is  
taken to be $0.254\,[{\rm GeV}^{-1}]$ for the charged decay and
$0.388\,[{\rm GeV}^{-1}]$ for the neutral 
decay~\cite{Eidelman:2004wy}.

Taking all of these into consideration, we construct the invariant
amplitudes as follows : 
\bee
i\mathcal{M}_{s}&=&-\frac{eg_{KN\Lambda^*}}{M_{K}}\bar{u}^{\mu}(p_{2},s_{2})
k_{2\mu}{\gamma}_{5} \frac{(\Slash{p}_{1}+M_{p})F_{1,c}+
\Slash{k}_{1}F_{1,s}}{q^{2}_{s}-M^{2}_{p}}
\Slash{\epsilon}u(p_{1},s_{1}),\nn\\&+&
\frac{e\kappa_{p}g_{KN\Lambda^*}}{2M_{p}M_{K}}
\bar{u}^{\mu}(p_{2},s_{2})k_{2\mu}{\gamma}_{5} 
\frac{(\Slash{q}_{s}+M_{p})F_{1,s}}{q^{2}_{s}-M^{2}_{p}}
\Slash{\epsilon}\Slash{k}_{1}u(p_{1},s_{1})\nn\\
i\mathcal{M}_{u}&=&-\frac{g_{KN\lambda}\kappa_{\Lambda^*}}{2M_{K}M_{\Lambda}}
\bar{u}_{\mu}(p_2)\Slash{k}_{1}\Slash{\epsilon}D^{\mu}_{\sigma}
\Theta^{\sigma\rho}k_{2\rho}\gamma_{5}u(p_1)F_{1,u},\nn\\ 
\mathcal{M}_{t}&=&\frac{2eg_{KN\Lambda^*}}{M_K}
\bar{u}^{\mu}(p_{2},s_{2})\frac{q_{t,\mu}k_{2}\cdot\epsilon}{q^{2}_{t}-M^{2}_{K}}
{\gamma}_{5}u(p_{1},s_{1})F_{1,c},\nn\\
i\mathcal{M}_{c}&=&\frac{eg_{KN\Lambda^*}}{M_K}\bar{u}^{\mu}(p_{2},s_{2})
\epsilon_{\mu}{\gamma}_{5}u(p_{1},s_{1})F_{1,c},\nn\\i\mathcal{M}_{v}&=&
\frac{-ig_{\gamma{K}K^*}g_{K^{*}NB}}{M_{K^{*}}(q^{2}_{t}-M^{2}_{K^*})}   
\bar{u}^{\mu}(p_{2},s_{2})\gamma_{\nu}\left(q^{\mu}_{t}g^{\nu\sigma}-
g^{\nu}_{t}q^{\mu\sigma}\right)\epsilon_{\rho\eta\xi\sigma}k^{\rho}_{1}
\epsilon^{\eta}k^{\xi}_{2}u(p_{1},s_{1})F_{1,v},
\label{amplitudes}  
\eee
where $\epsilon$ and ${u}^{\mu}$ are the photon polarization vector
and the RS vector-spinor which is defined as follows:
\bee
u^{\mu}(p_{2},s_2)=\sum_{\lambda,s}\left\langle1\lambda\frac{1}{2}s\left|
\frac{3}{2}s_{2}\right\rangle\right. e^{\mu}(p_{2},\lambda)u(p_{2},s)
\eee 
with the Clebsh-Gordon coefficient
$\langle 1\lambda\frac{1}{2}s|\frac{3}{2}s_{2}\rangle$.  $D_{\mu\nu}$
stands for the spin-3/2 propagator: 
\bee
D_{\mu\nu}=-\frac{\Slash{q}+M_{\Lambda^*}}{q^2-M^2_{\Lambda^*}}
\left[g_{\mu\nu}-\frac{1}{3}\gamma_{\mu}\gamma_{\nu}-
\frac{2}{3M^{2}_{\Lambda^*}}q_{\mu}q_{\nu}+\frac{q_{\mu}\gamma_{\nu}+
q_{\mu}\gamma_{\nu}}{3M_{\Lambda^*}}\right].
\eee
In Eq.~(\ref{amplitudes}), we have shown how the four-dimensional form
factor is inserted in such way that gauge-invariance is preserved.  As
suggested in Ref.~\cite{Haberzettl:1998eq,Davidson:2001qs}, we
adopt the following parameterization for the four-dimensional form
factors: 
\bee
F_{1,x}(q^2)&=&\frac{\Lambda_1^4}{\Lambda_1^4+(x-M^2_x)^2},\,\,x=s,t,u,v\nn\\
F_{1,c}&=&F_{1,s}+F_{1,t}-F_{1,s}F_{1,t}. 
\label{formfactor1}
\eee
The form of $F_{1,c}$ is chosen such that the on-shell values of the
coupling constants are reproduced. 

We consider another type of the form factor with the three-momentum
cutoff, which is parameterized as follows:
\be
F_2
(|\vec{k}_1|,|\vec{k}_2|)=\left(\frac{\Lambda_2^2}{\Lambda_2^2+|\vec{k}_1|^2}
\right)\left(\frac{\Lambda_2^2-P^2_{KN\Lambda^*}}{\Lambda_2^2+|\vec{k}_2|^2}\right),  
\label{formfactor2}
\ee
where $k_1$ and $k_2$ denote the momenta of the initial photon and 
final kaon, respectively.  We will multiply all the
amplitudes $\mathcal{M}_{s,t,u,c,v}$ by the form factor $F_2$ to
maintain gauge-invariance.  In order to satisfy the normalization condition for
the $KN\Lambda^*$ vertex, we set $P_{KN\Lambda^*} = 238$ MeV,
considering the decay process $\Lambda^*\to \bar{K}N$.  The cutoff
masses $\Lambda_1$ and $\Lambda_2$ will be adjusted to produce the
data of the total cross section $\sigma_{\gamma p\to
  K^+\Lambda^*}$. 

\section{Numerical results}
\subsection{$\gamma N\to K\Lambda^*$ without the form factors}
\begin{figure}[tbh] 
\begin{tabular}{cc}
\resizebox{8cm}{6cm}{\includegraphics{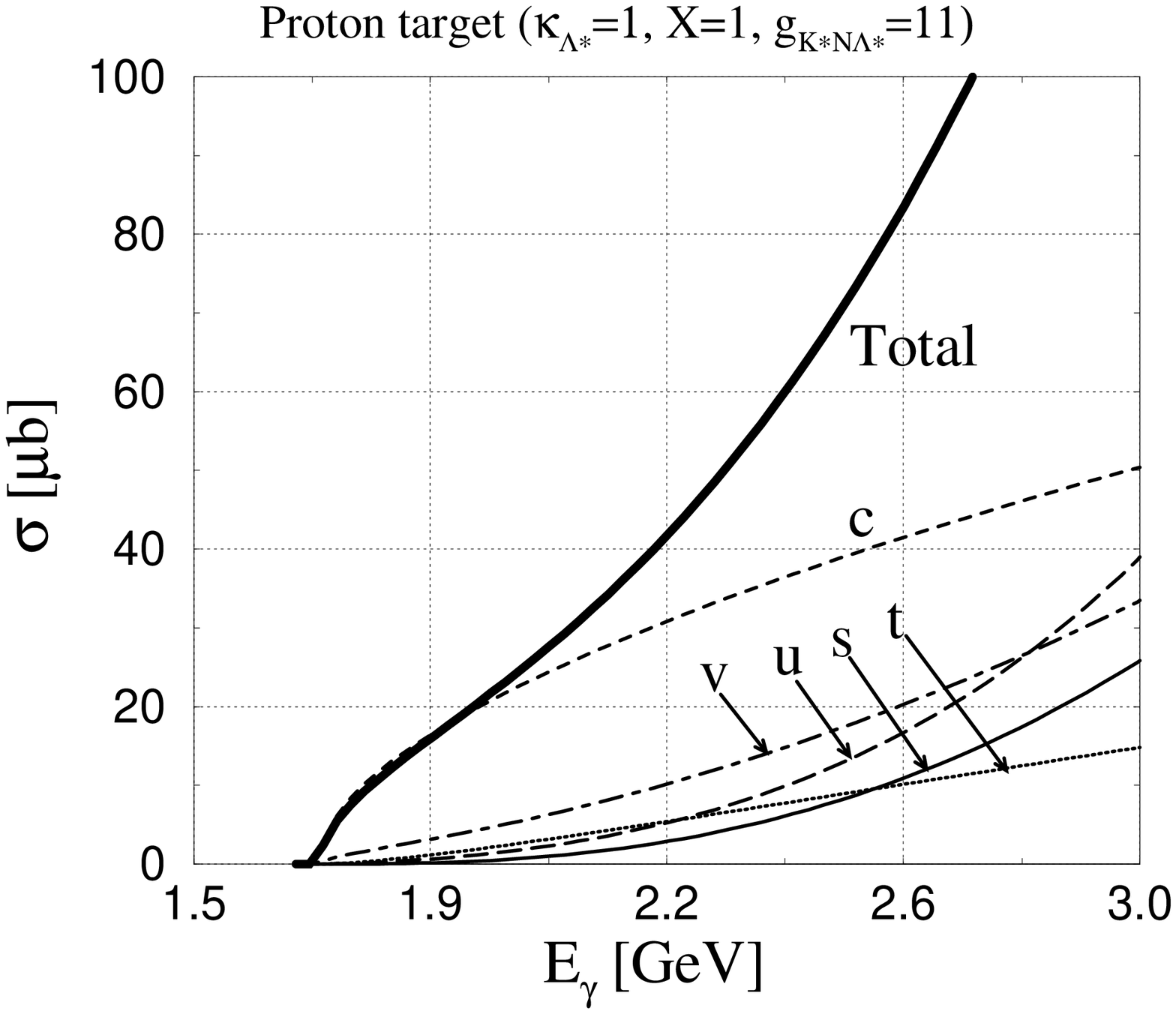}}
\resizebox{8cm}{6cm}{\includegraphics{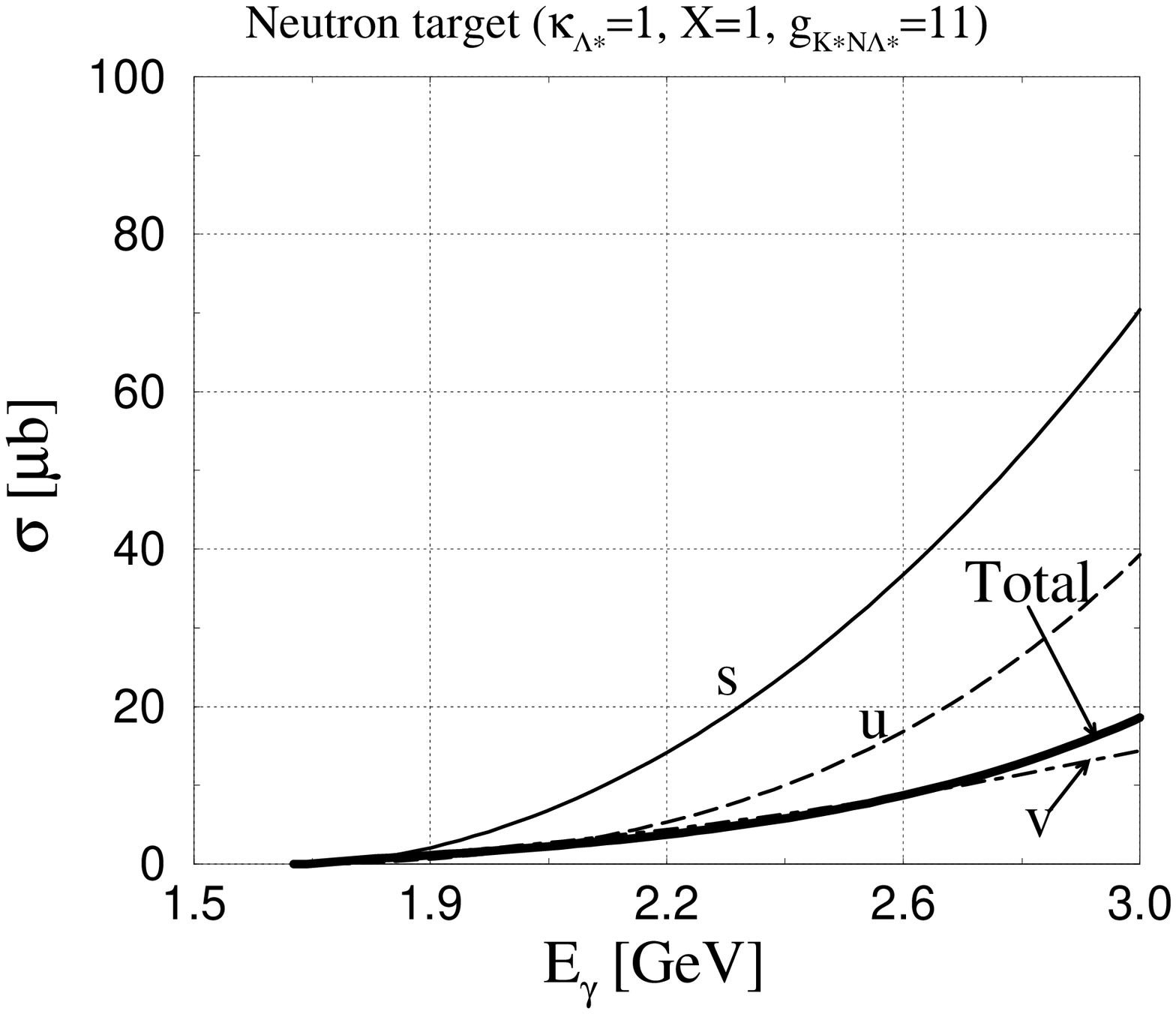}}
\end{tabular}
\caption{Each contribution of various channels to the total cross sections
without form factors.  In the left panel, the total cross section for
the proton target is depicted, while in the right panel that for the
neutron one is drawn.  We choose the parameters as follows:
$(\kappa_{\Lambda^*},X)=(1,1)$ and $g_{KN\Lambda^*}=+11$.}
\label{fig1}
\end{figure} 
We first consider the numerical results for the total cross section of
the $\gamma N\to K\Lambda^*$ process without form factors to examine
the contributions of various channels which are depicted as functions
of the incident photon energy $E_\gamma$ in Fig.~\ref{fig1}.  Here, we
choose the unknown parameters as follows: $\kappa_{\Lambda^*}=1.0$,
$X=1$ and $g_{K^*N\Lambda^*}=g_{KN\Lambda^*}=+11$.  The parameter
dependence will be discussed later.  The contact term,
i.e. $c$--channel is dominant over all other channels for the proton
target, as shown in the left panel of Fig.~\ref{fig1}.  Near the
threshold, in particular, the $c$-- and $v$--channels are
characterized by the energy dependence of the $s$--wave type,
i.e. $\sigma\sim (E_{\gamma}-E_{\rm   th})^{1/2}$, where $E_{\rm th}$
stands for the threshold energy, although the magnitude of the
$v$--channel is much smaller than that of the $c$--channel.  On the
other hand, the $s$--, $u$--, and $t$--channels are governed by the
$p$-wave, due to which their contributions turn out to be much smaller
than those of the $c$-- and $v$--ones in the vicinity of the
threshold.  

In the case of the neutron target, as depicted in the right panel of
Fig.~\ref{fig1}, the contact term is absent, which makes  
the $s$--channel the largest.  Moreover, we also find a destructive
interference between the $s$--, $u$-- and $v$--channels, so that the
total cross section for the neutron target turns out to be much
smaller than that for the proton one.  Although it is not shown here,
we verified that forward scattering is enhanced both in the proton and
neutron targets when the form factors are turned off.
\subsection{$\gamma N\to K\Lambda^*$ with the form factor $F_1$}
We are now in a position to introduce the form factor $F_1$ defined as
in Eq.~(\ref{formfactor1}). 
\begin{figure}[tbh]
\begin{center}
\resizebox{8cm}{6cm}{\includegraphics{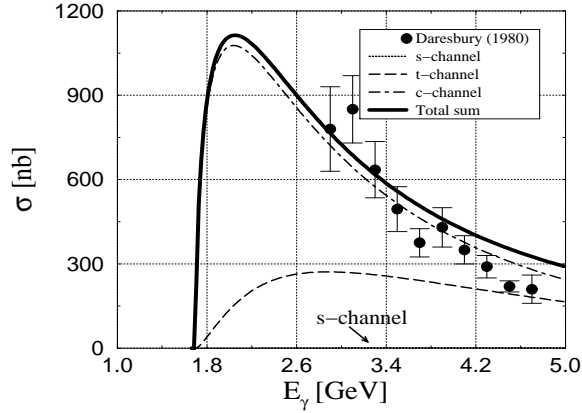}}
\end{center}
\caption{The total cross sections for the proton target with the
form factor $F_1$.  The $s$--, $t$-- and
$c$--channel contributions are drawn separately.  The experimental
data are taken from Ref.~\cite{Barber:1980zv}} 
\label{fig2}
\end{figure} 
In Fig.~\ref{fig2}, the $s$--, $t$-- and $c$--channel contributions to
the total cross section for the proton target are drawn, separately.
Note that they do not contain the parameters such as 
$\kappa_{\Lambda^*}$, $X$ and $g_{K^*N\Lambda^*}$.  The experimental
data are taken from Ref.~\cite{Barber:1980zv} in the range of the
photon energy: $2.8$ GeV $<\, E_{\gamma}\,<$  $4.8$ GeV.  The cutoff
parameter is fixed to reproduce the experimental data.  $\Lambda_1=750$
MeV is our best value.  However, the results at higher energies should
be taken cautiously, since the Born approximation is reliable only in 
the low-energy region near the threshold.  In fact, we have found
that the total cross section depends much on the parameters such as
$\kappa_{\Lambda^*}$ and $X$ beyond $E_{\gamma}\gsim3$ GeV, whereas
it turns out that its dependence on those parameters is rather 
weak when $E_{\gamma}\lsim3$ GeV~\cite{nam1}.   Therefore, we
focus most of our discussion below $E_{\gamma}\lsim3$ GeV, where the
Born approximation is expected to be reliable.  It is interesting to
observe that the size and energy dependence 
of the total cross section of the $\Lambda^*(1520)$ production
are similar to those of the production of the ground state
$\Lambda(1116)$~\cite{Boyarski:1970yc,Barber:1980zv}.    
 
As shown in Fig.~\ref{fig2}, the $c$--channel is the most dominant
contribution without which one can never reproduce the data for the
proton target.  On the other hand, the $s$--channel contribution is
almost negligible and the $t$--channel is marginal.  

\begin{figure}[tbh]
\begin{tabular}{cc}
\resizebox{8cm}{5.5cm}{\includegraphics{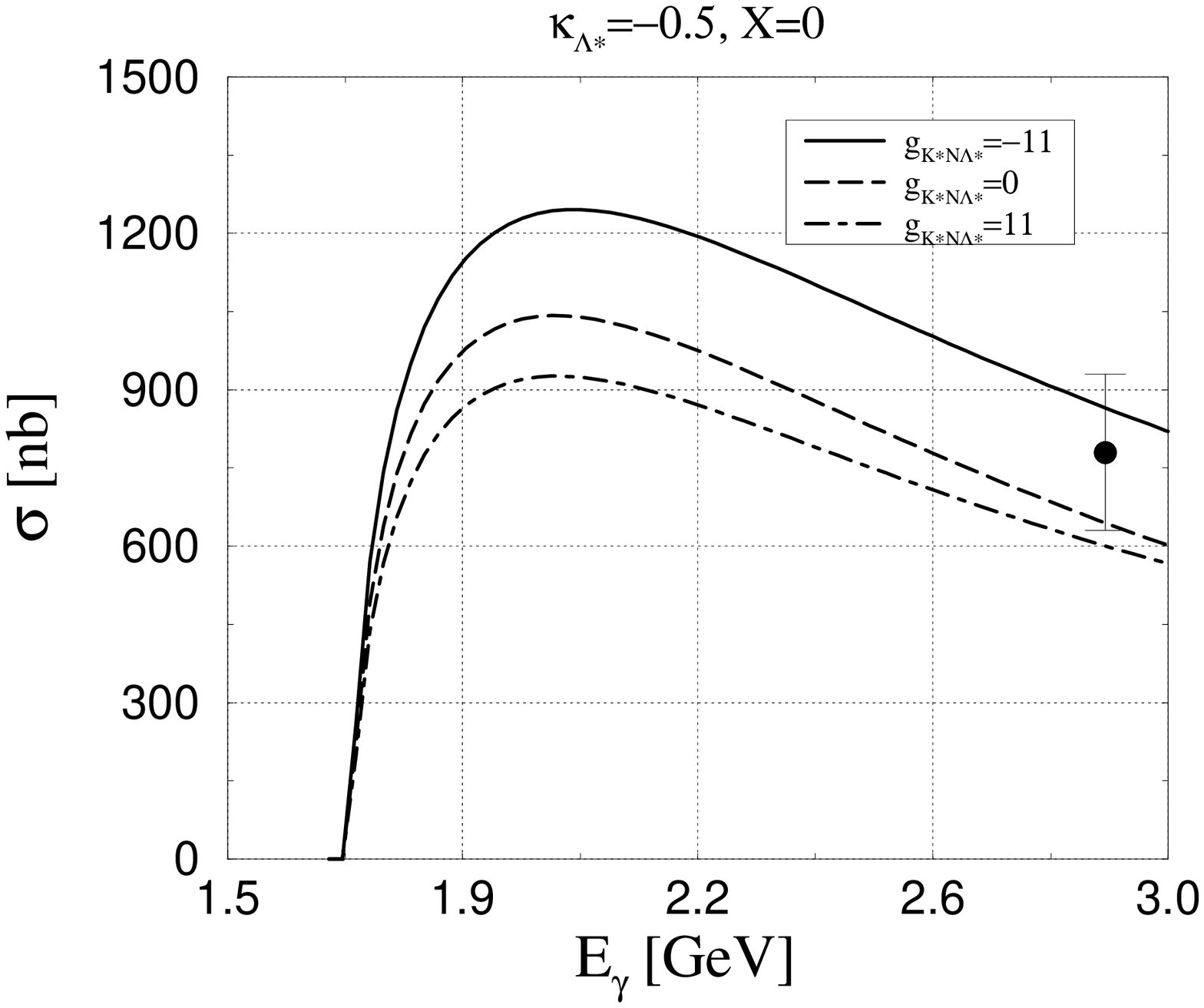}}
\resizebox{8cm}{5.5cm}{\includegraphics{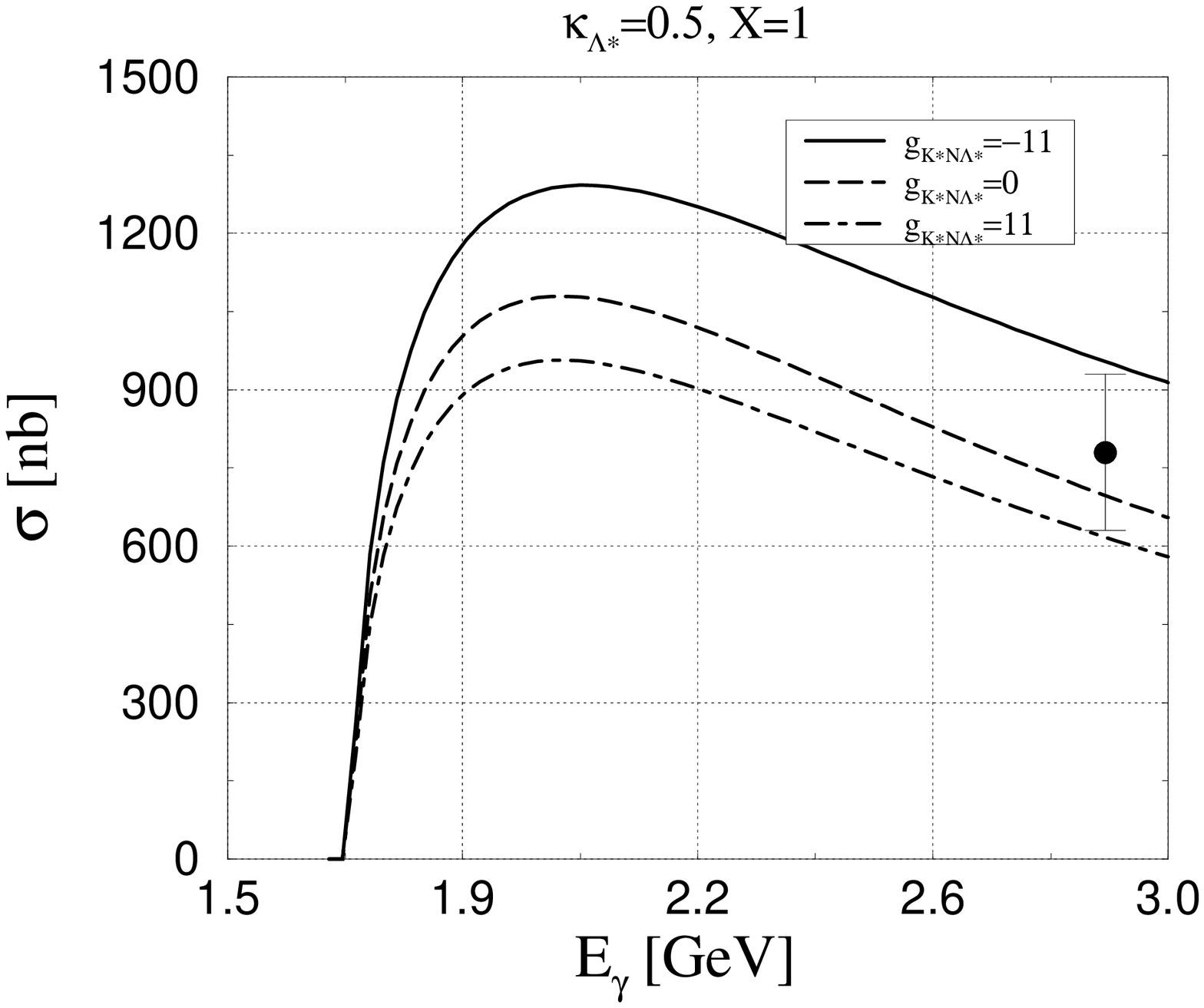}}
\end{tabular}
\caption{The total cross sections for the proton target with the
form factor $F_1$. We choose $(\kappa_{\Lambda^*},X) = (-0.5,0)$ and
$(0.5,1)$ in order to see the parameter dependence. We choose three
different values of the coupling constant $g_{K^*N\Lambda^*}=0$ and
$\pm11$.}   
\label{fig3}
\end{figure}

Figure~\ref{fig3} depicts the total cross sections with the coupling
constant $g_{K^*N\Lambda^*}$ varied.  We basically use the relation
$g_{K^*N\Lambda^*}=\pm|g_{KN\Lambda^*}|$, i.e. $g_{K^*N\Lambda^*}=\pm
11$.       
The total cross sections are rather insensitive to its values.  

These two results are also compared for the two different parameter
sets, i.e. $(\kappa_{\Lambda^*},X)=(-0.5,0)$ and $(0.5,1)$.  As
discussed previously, the results do not depend 
much on these parameters at $E_{\gamma}\lsim3$ GeV.  In
the quark model, it is found that the anomalous magnetic moment
$\kappa_{\Lambda^*}$ turns out to vanish in pure $SU(3)$
symmetry.  Taking into account explicit $SU(3)$ 
symmetry breaking, we expect that the values of $\kappa_{\Lambda^*}$
may lie in the range of $|\kappa_{\Lambda^*}|<0.5$.  However, we find from
Fig.~\ref{fig3} that the dependence on $\kappa_{\Lambda^*}$
within this range is rather small.  Therefore, these two parameters
$\kappa_{\Lambda^*}$ and $X$ can be safely set to be zero,
i.e. $\kappa_{\Lambda^*}=X=0$.  

\begin{figure}[tbh]
\begin{tabular}{c}
\resizebox{8cm}{5.5cm}{\includegraphics{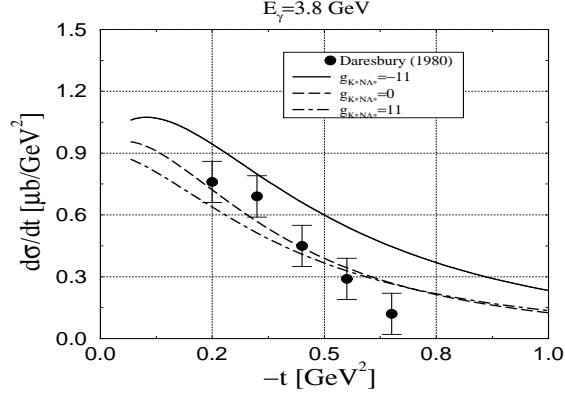}}
\end{tabular}
\caption{The $t$--dependence for the proton target at
$E_{\gamma}=3.8$ GeV. We choose  $(\kappa_{\Lambda^*},X) = (0,0)$. The
experimental data are taken from Ref.~\cite{Barber:1980zv}}  
\label{fig4}
\end{figure} 

\begin{figure}[tbh]
\begin{tabular}{ccc}
\resizebox{5.5cm}{5.5cm}{\includegraphics{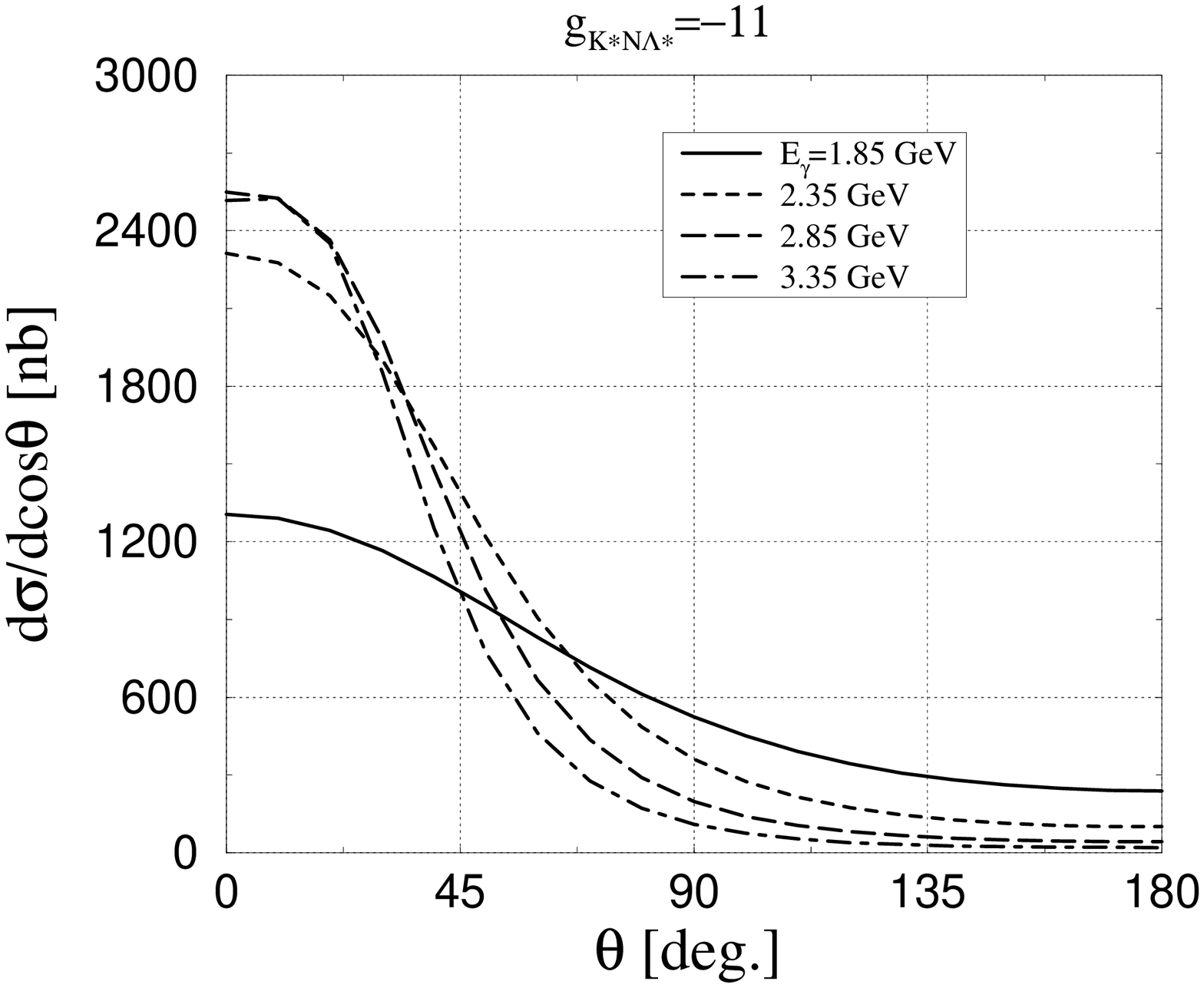}}
\resizebox{5.5cm}{5.5cm}{\includegraphics{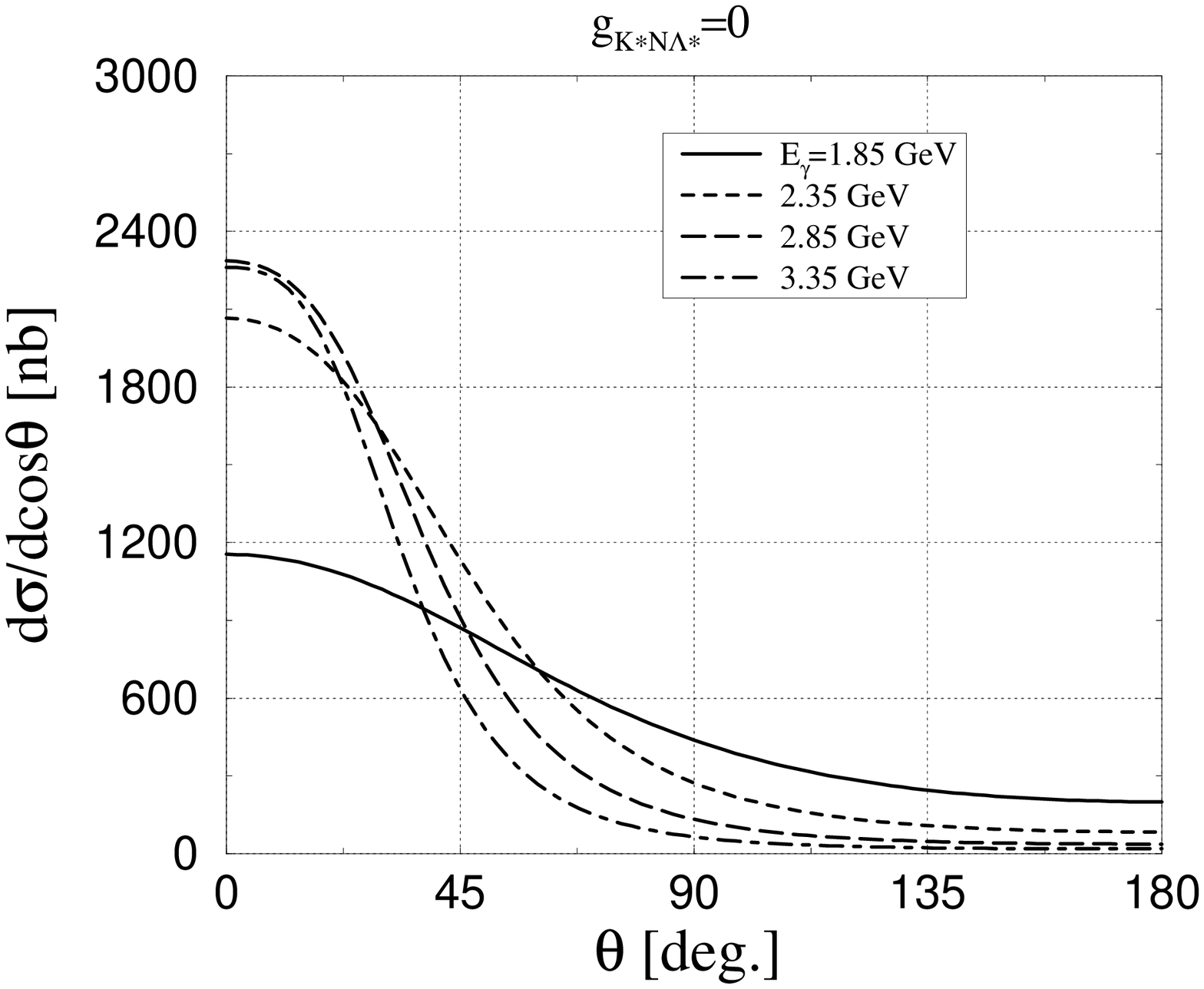}}
\resizebox{5.5cm}{5.5cm}{\includegraphics{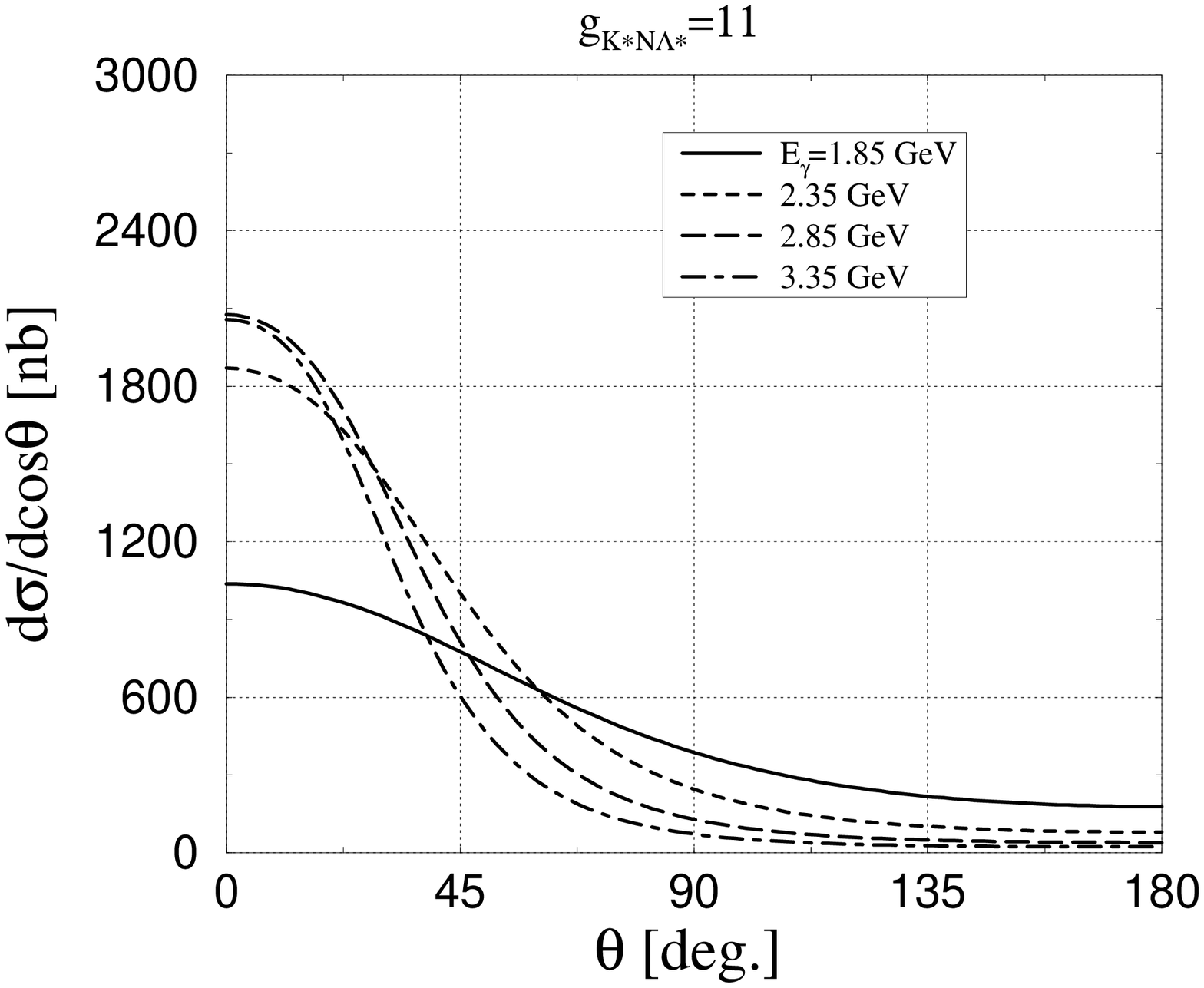}}
\end{tabular}
\caption{The differential cross sections for the proton target with the
form factor $F_1$.  Several photon energies are taken into
account. We choose  $(\kappa_{\Lambda^*},X) = (0,0)$.}  
\label{fig5}
\end{figure}

In Fig.~\ref{fig4}, we depict the dependence on the momentum transfer,
${d\sigma}/{dt}$ ($t$--dependence) at 
$E_{\gamma}=3.8$ GeV which is the average energy of the Daresbury
experiment  ($2.8<E_{\gamma}<4.8$ GeV)~\cite{Barber:1980zv}.  The
figure indicates that the present work is in good agreement with the
data.  In Fig.~\ref{fig5}, we also demonstrate the angular
dependence.  Here, $\theta$ is the angle between the incident photon
and the outgoing kaon in the center of mass system.  Each panel draws
the differential cross sections ${d\sigma}/{d(\cos\theta)}$ with
$g_{K^*N\Lambda^*}$ varied.  We observe that $K^*$--exchange does not
contribute much to the differential cross sections as in the case of
the total cross sections (see also Fig.~\ref{fig3}). 

\begin{figure}[tbh]
\begin{tabular}{cc}
\resizebox{8cm}{5.5cm}{\includegraphics{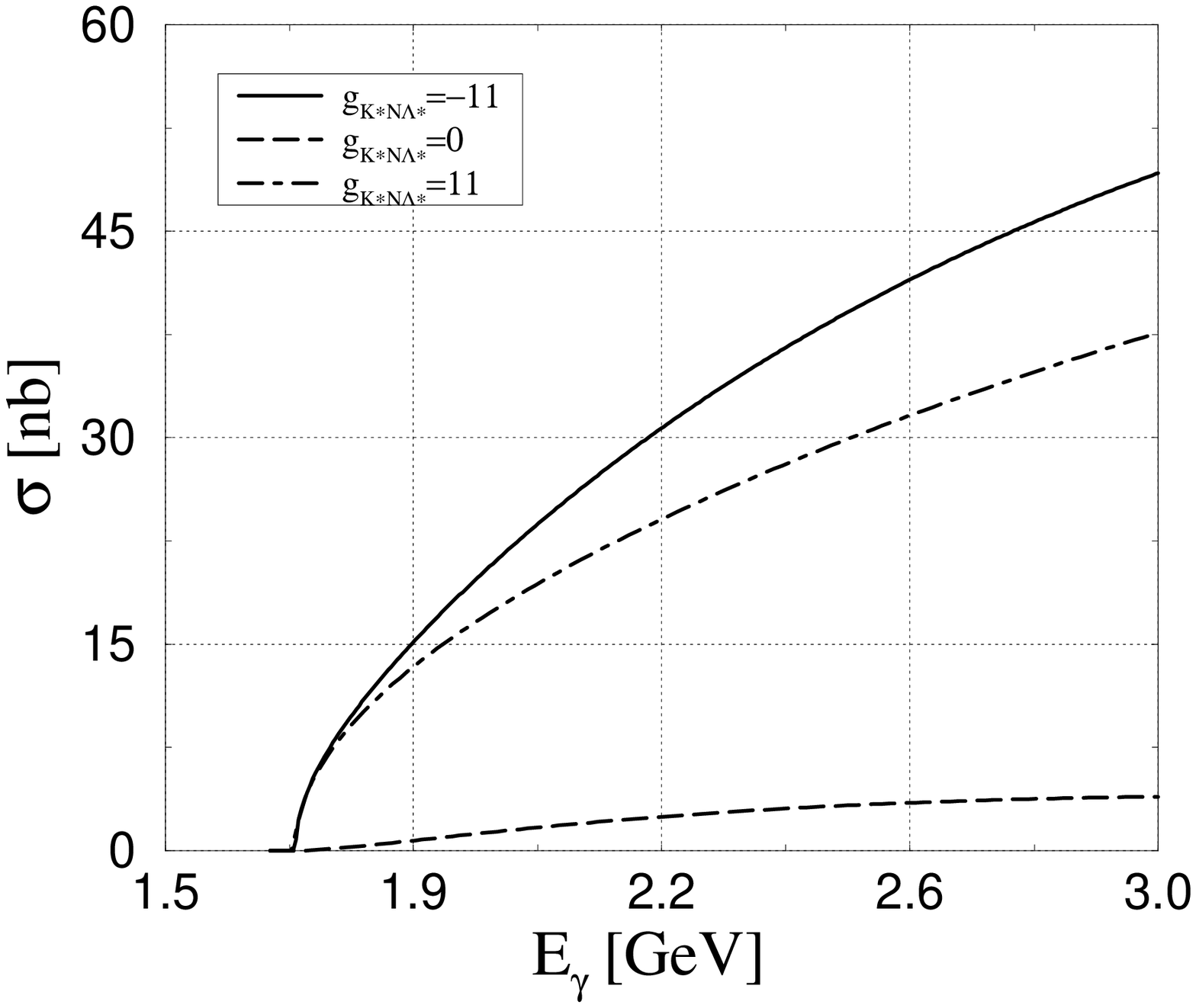}}
\resizebox{8cm}{5.5cm}{\includegraphics{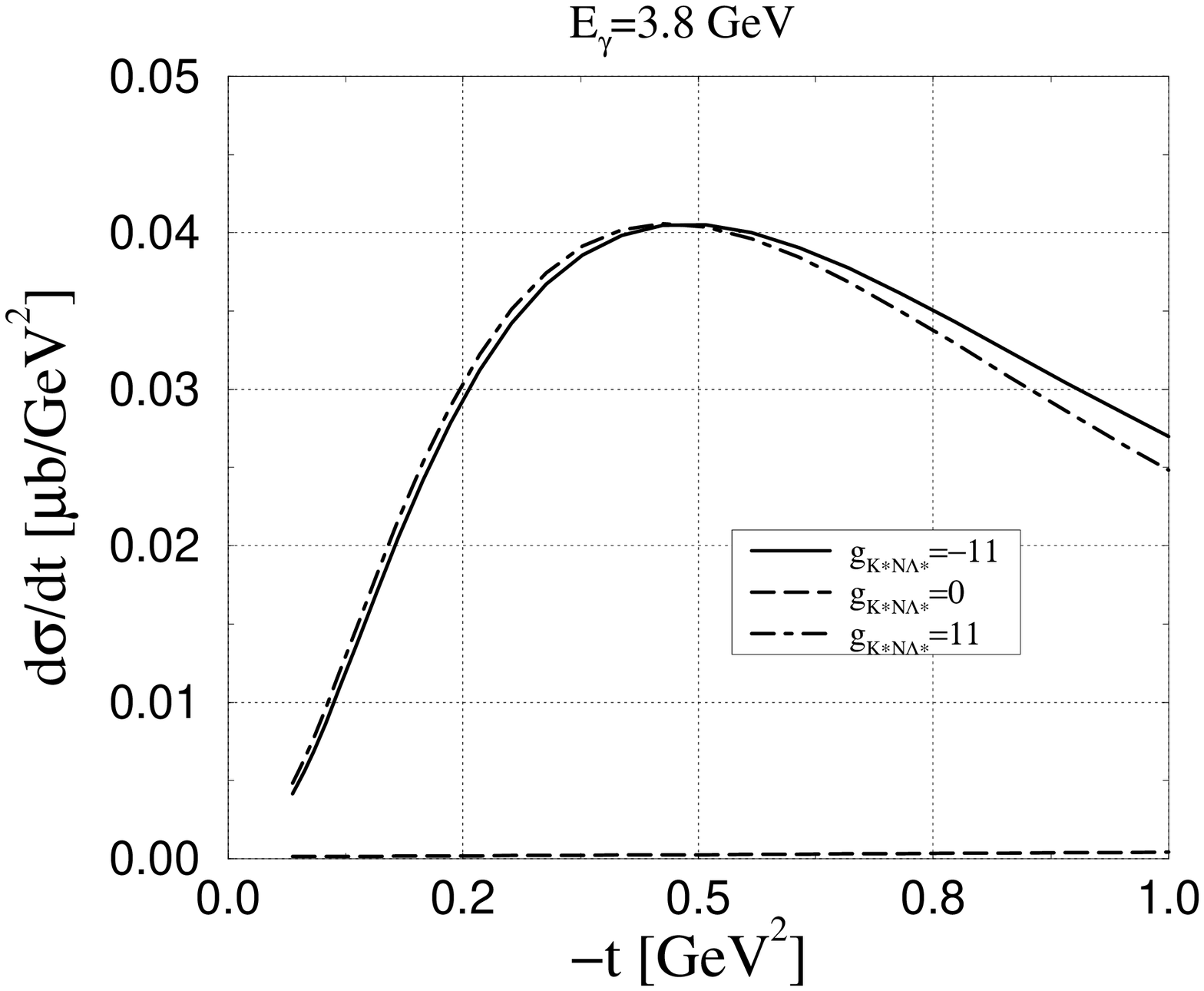}}
\end{tabular}
\caption{In the left panel, the total cross sections are depicted for
  the neutron target with the form factor $F_1$, while in the right
  panel the $t$--dependence is drawn at $E_{\gamma}=3.8$ GeV.  We choose
  $(\kappa_{\Lambda^*},X) = (0,0)$ and three different values of the 
coupling constants, i.e. $g_{K^*N\Lambda^*}=0$ and $\pm 11$.} 
\label{fig6}
\end{figure}

Figure~\ref{fig6} predicts the total cross section for the neutron
target and $d\sigma/dt$ at $E_\gamma =3.8$ GeV with the form factor $F_1$
being employed.  In this case, the contact term is absent, since the
process $\gamma n\rightarrow K^0\Lambda^*$ is the neutral one (see 
Eq.(\ref{amplitudes})).  Its absence causes the total cross section to
become much smaller than that for the proton target.  The left panel
of Fig.~\ref{fig6} depicts the total cross sections with the 
three different values of the coupling constant $g_{K^*N\Lambda^*}$.
$K^*$--exchange plays a significant role in describing the $\Lambda^*$
production off the neutron.  Furthermore, the total cross section is
proportional to $\sim(E_{\gamma}-E_{\rm th})^{1/2}$ by 
$K^*$--exchange.  When $K^*$ exchange is switched off,
i.e. $g_{K^*N\Lambda^*}=0$, the total cross section is strongly
suppressed and the $p$--wave is found to be dominant, so that its 
energy dependence is changed to be proportional to $(E_{\gamma}-E_{\rm 
  th})^{3/2}$ as shown in Fig.~\ref{fig6}.   

In the right panel of Fig.~\ref{fig6}, $d\sigma /dt$ is drawn at
$E_{\gamma}=3.8$ GeV.  It is natural that the $t$-dependence for the
neutron target be very different from that for the proton, since
dominant diagrams are different for each case.         

\begin{figure}[tbh]
\begin{tabular}{ccc}
\resizebox{5.5cm}{5.5cm}{\includegraphics{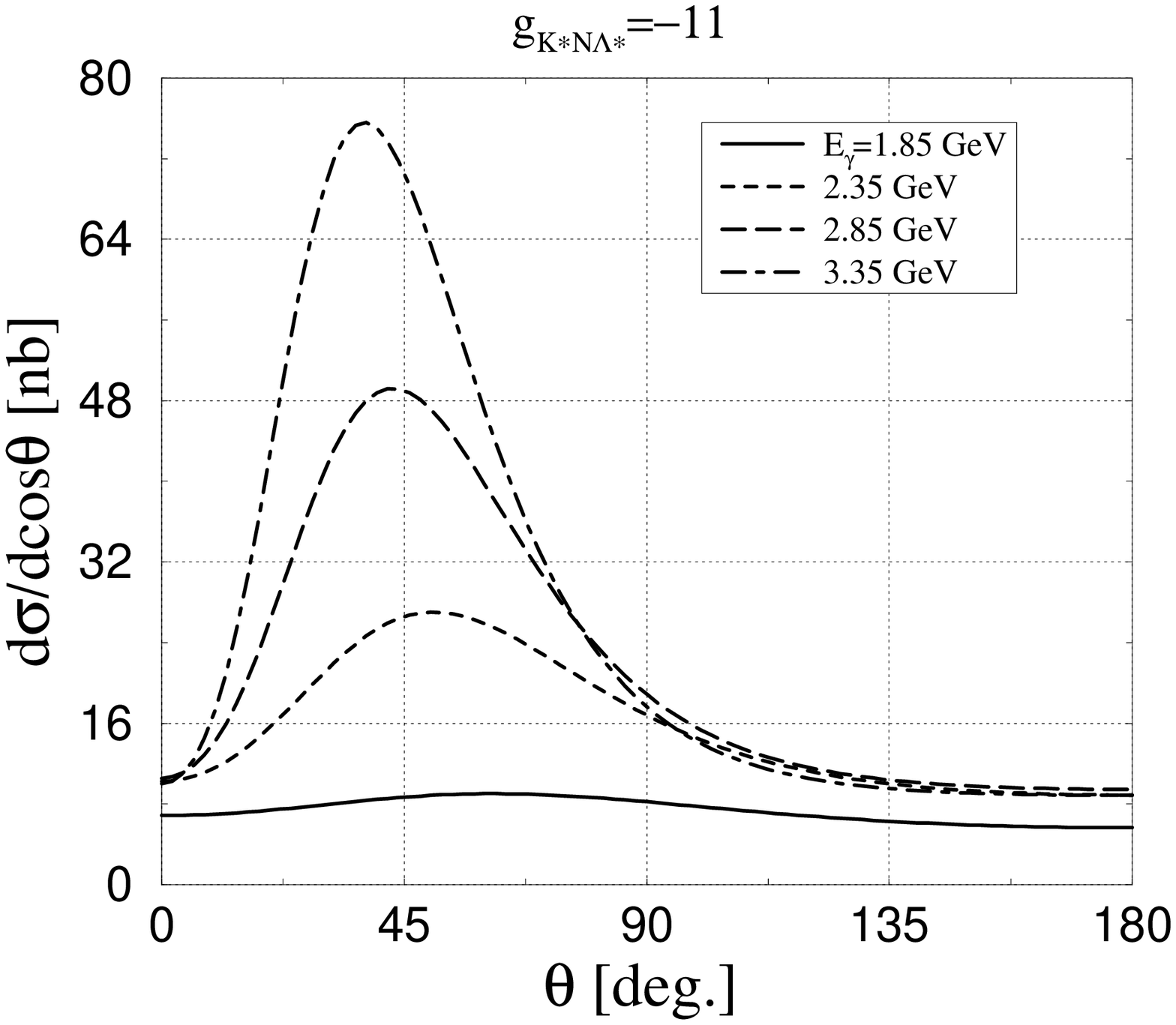}}
\resizebox{5.5cm}{5.5cm}{\includegraphics{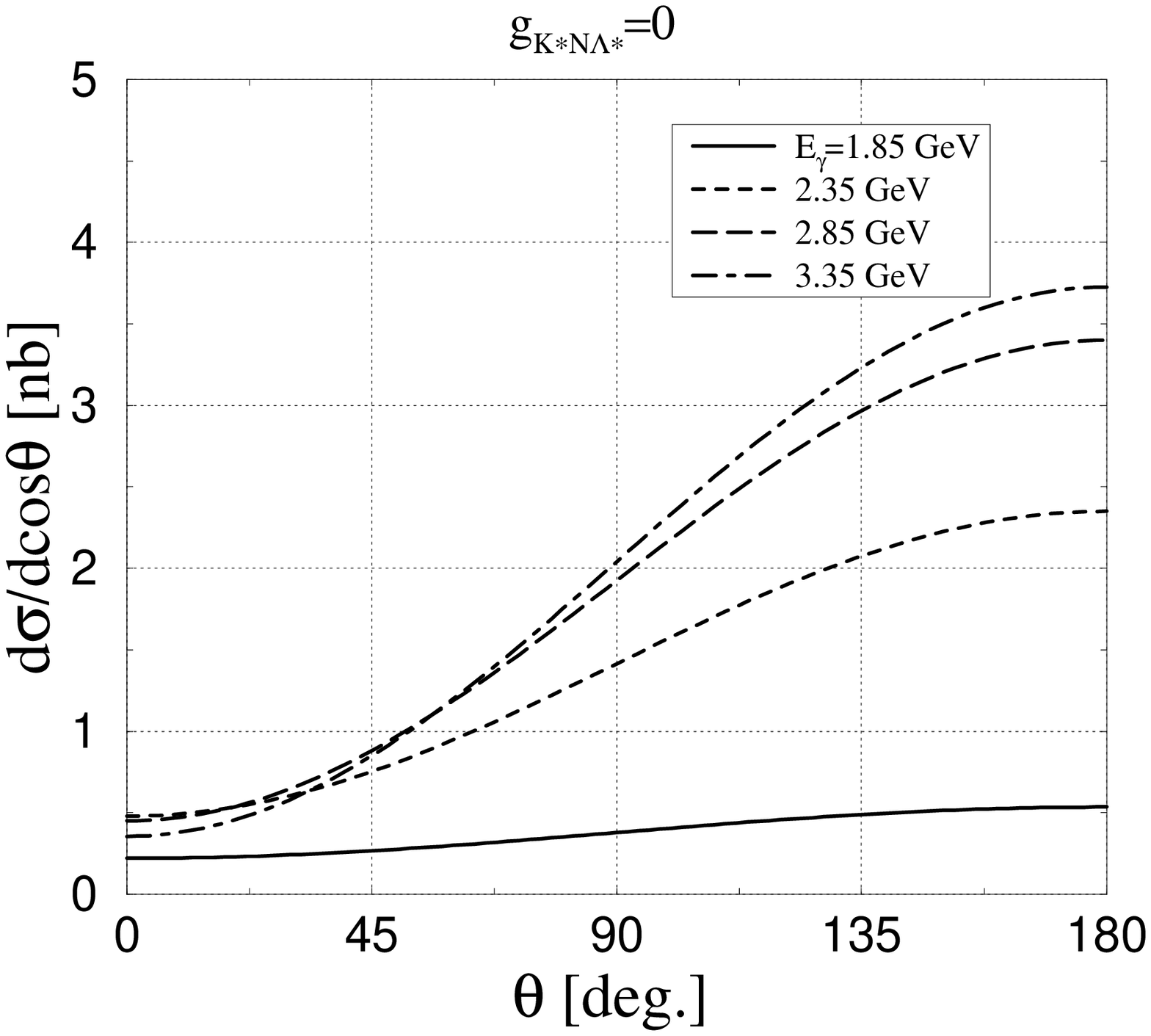}}
\resizebox{5.5cm}{5.5cm}{\includegraphics{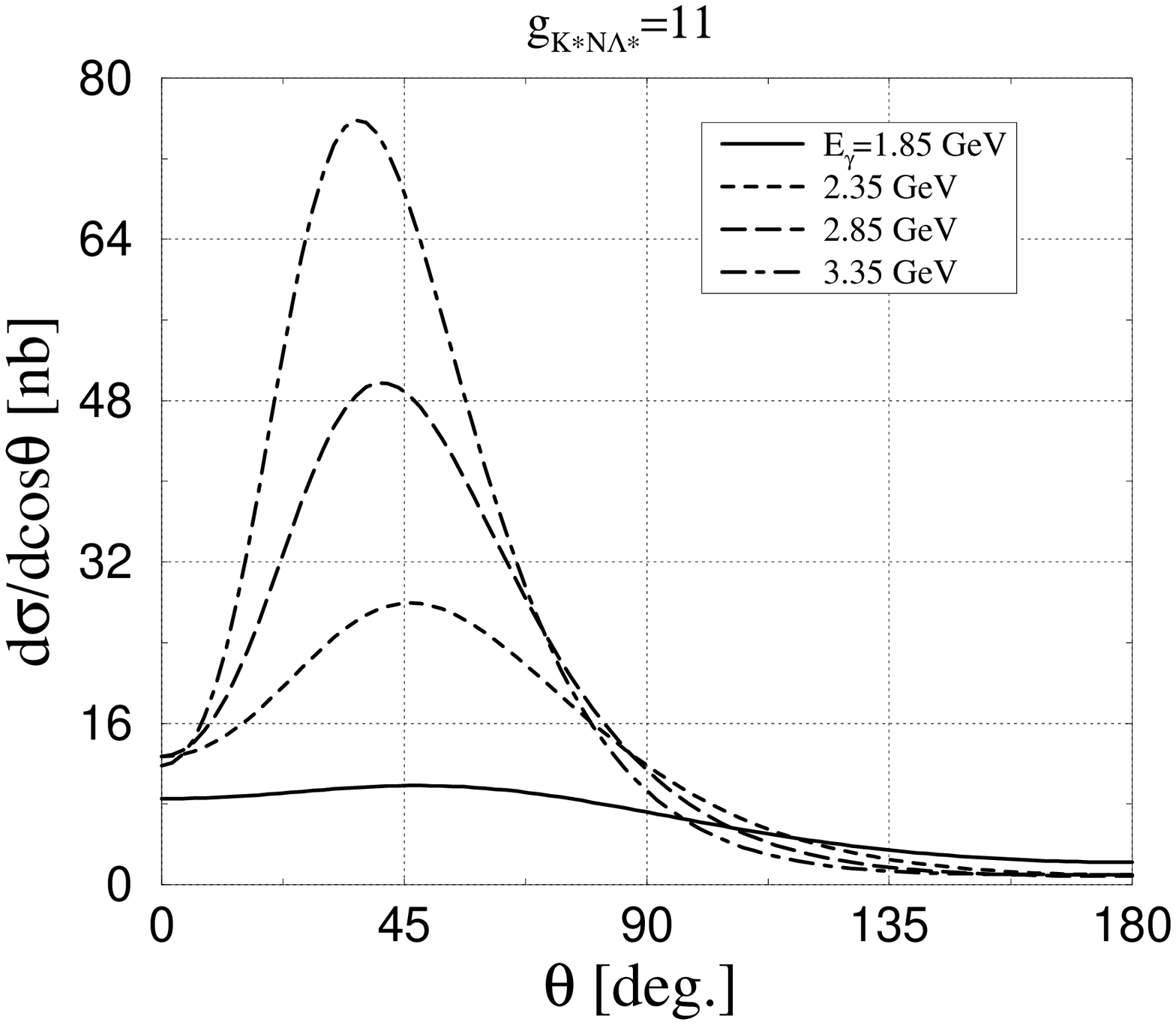}}
\end{tabular}
\caption{The differential cross sections for the neutron target with the
form factor $F_1$.  Several photon energies are taken into
account. We choose  $(\kappa_{\Lambda^*},X) = (0,0)$.}
\label{fig7}
\end{figure}

Figure~\ref{fig7} presents the differential cross sections for the
neutron target with three different values of $g_{K^*N\Lambda^*}$.
While the sign of the coupling constant does not change the results,
its absolute value seems to be of great importance.  For example,
$K^*$--exchange being included with $g_{K^*N\Lambda^*}=\pm 11$,   
the differential cross sections are enhanced around $45^{\circ}$.
Note that the sign of $g_{K^*N\Lambda^*}$ is not 
important. The bump around $45^{\circ}$ is a typical behavior when 
we introduce the form factor like $F_1$ in the $t$--channel.  

\subsection{$\gamma N\to K\Lambda^*$ with the form factor $F_2$}
When the form factor $F_2$ defined in Eq.(\ref{formfactor2}) is used, 
the results turn out to be quite different from those with $F_1$.  We 
determine the cutoff mass $\Lambda_2=650$ MeV by fitting the total
cross section to the  experimental data around $E_{\gamma}\sim 3$ GeV
from Ref.~\cite{Barber:1980zv}.  As in the previous case,  
we fix two parameters $(\kappa_{\Lambda^*},X)=(0,0)$, so that
the relevant contributions are from the $s$--, $t$--, $c$-- channels 
and $K^*$--exchange. 

\begin{figure}[tbh]
\resizebox{8cm}{5.5cm}{\includegraphics{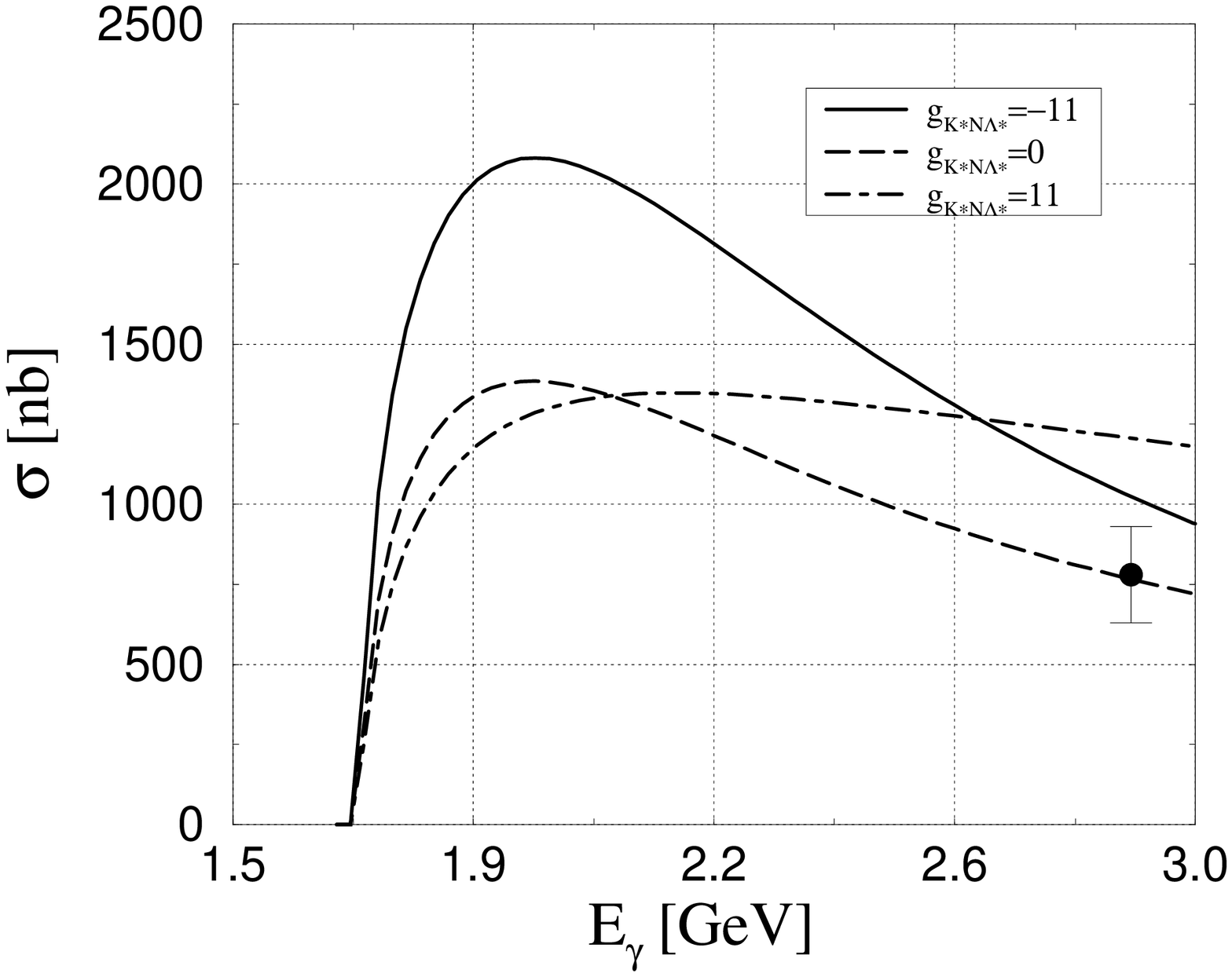}}
\resizebox{8cm}{5.5cm}{\includegraphics{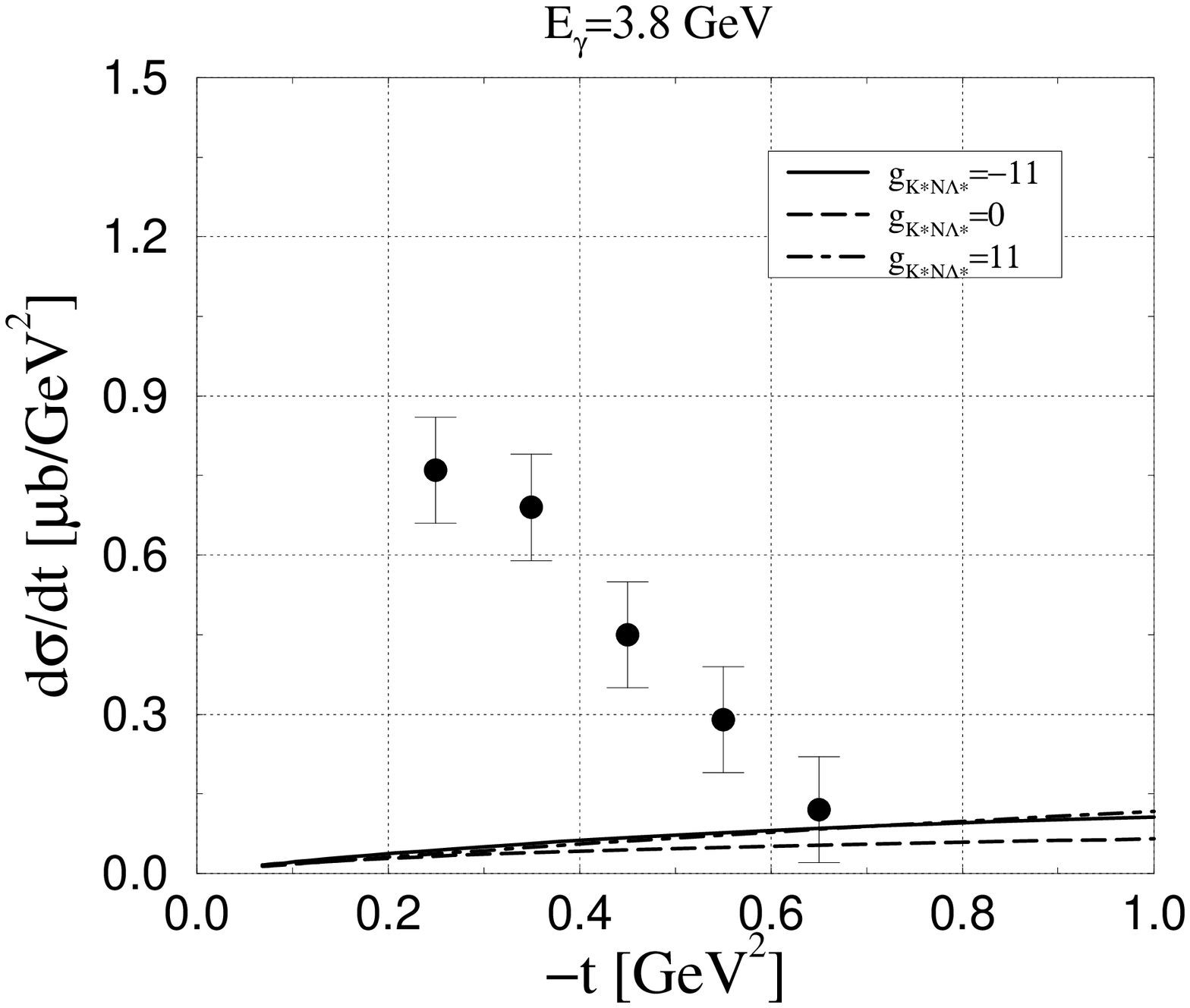}}
\caption{In the left panel, the total cross sections are depicted for
the proton target with the form factor $F_1$, while in the right
panel the $t$--dependence is drawn for the proton target at
$E_{\gamma}=3.8$ GeV.  We choose   $(\kappa_{\Lambda^*},X) = (0,0)$
and three different values of the coupling constants,
i.e. $g_{K^*N\Lambda^*}=0$ and $\pm 11$.}
\label{fig8}
\end{figure}

In the left panel of Fig~\ref{fig8}, we show the total cross sections
as functions of the incident photon energy $E_{\gamma}$ for the proton  
target.  While the energy dependence looks similar to that
with the form factor $F_1$ as drawn in Fig.~\ref{fig3}, the magnitude is 
quite larger than that.  Moreover, the energy dependence of $K^*$
exchange is changed by replacing the form factor $F_1$ by the $F_2$ as
shown in Fig~\ref{fig8}.  This can be understood by comparing the $F_2$
with the $F_1$, defined in Eqs.(\ref{formfactor1}, \ref{formfactor2}),
respectively.  While the form factor $F_2$ has an overall energy
dependence, the $F_1$ does not.

We plot the $t$--dependence for the proton target in the 
right panel of Fig.~\ref{fig8}, using the form factor $F_2$.  The
curves show quite different $t$--dependence from those with the $F_1$
(Fig.~\ref{fig4}).  Thus, the results deviate from
the data when using the $F_2$, though we have obtained the reasonable size
and energy dependence of the total cross sections as shown in the left   
panel of Fig.~\ref{fig8}. 
\begin{figure}[tbh]
\begin{tabular}{ccc}
\resizebox{5.5cm}{5.5cm}{\includegraphics{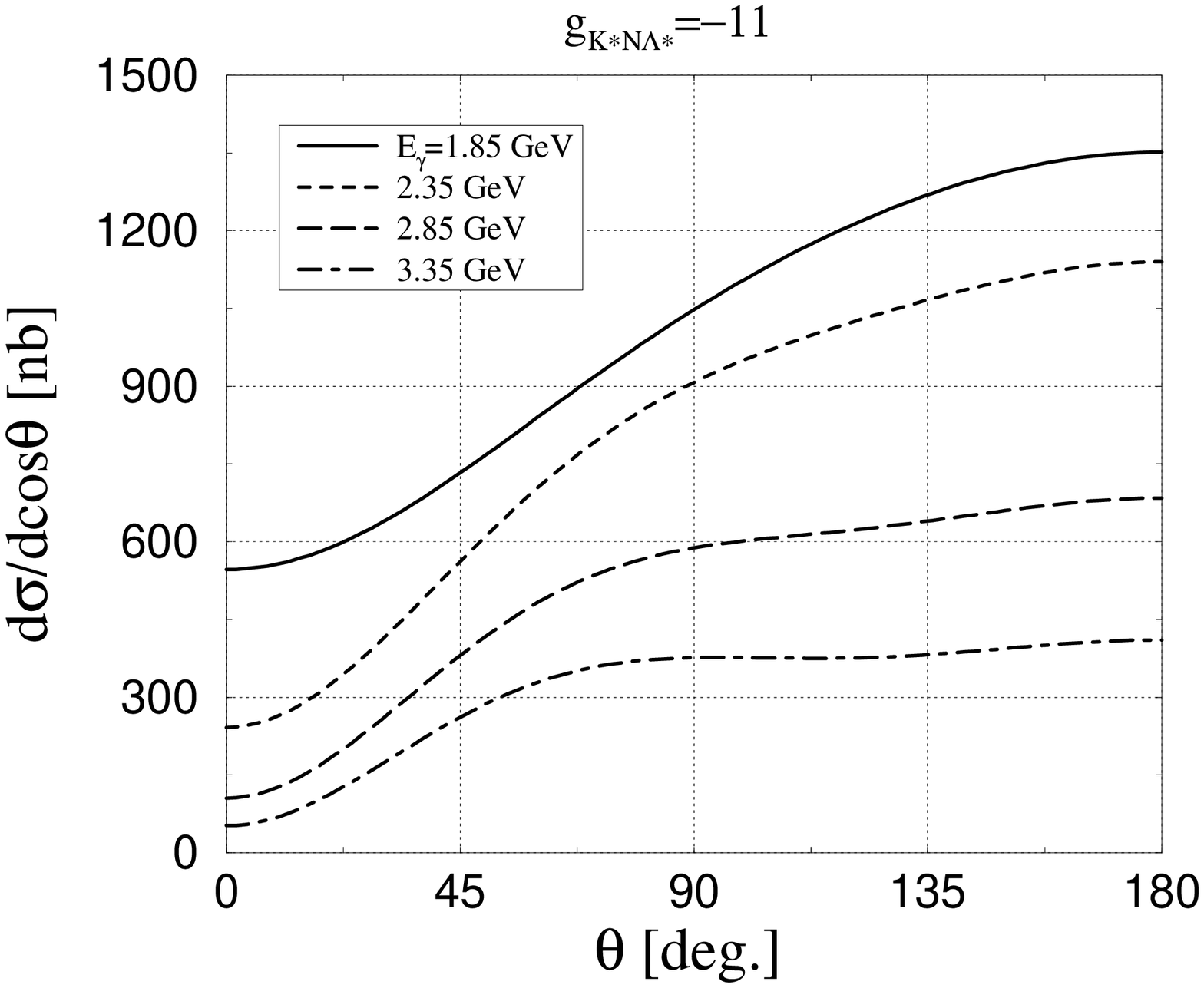}}
\resizebox{5.5cm}{5.5cm}{\includegraphics{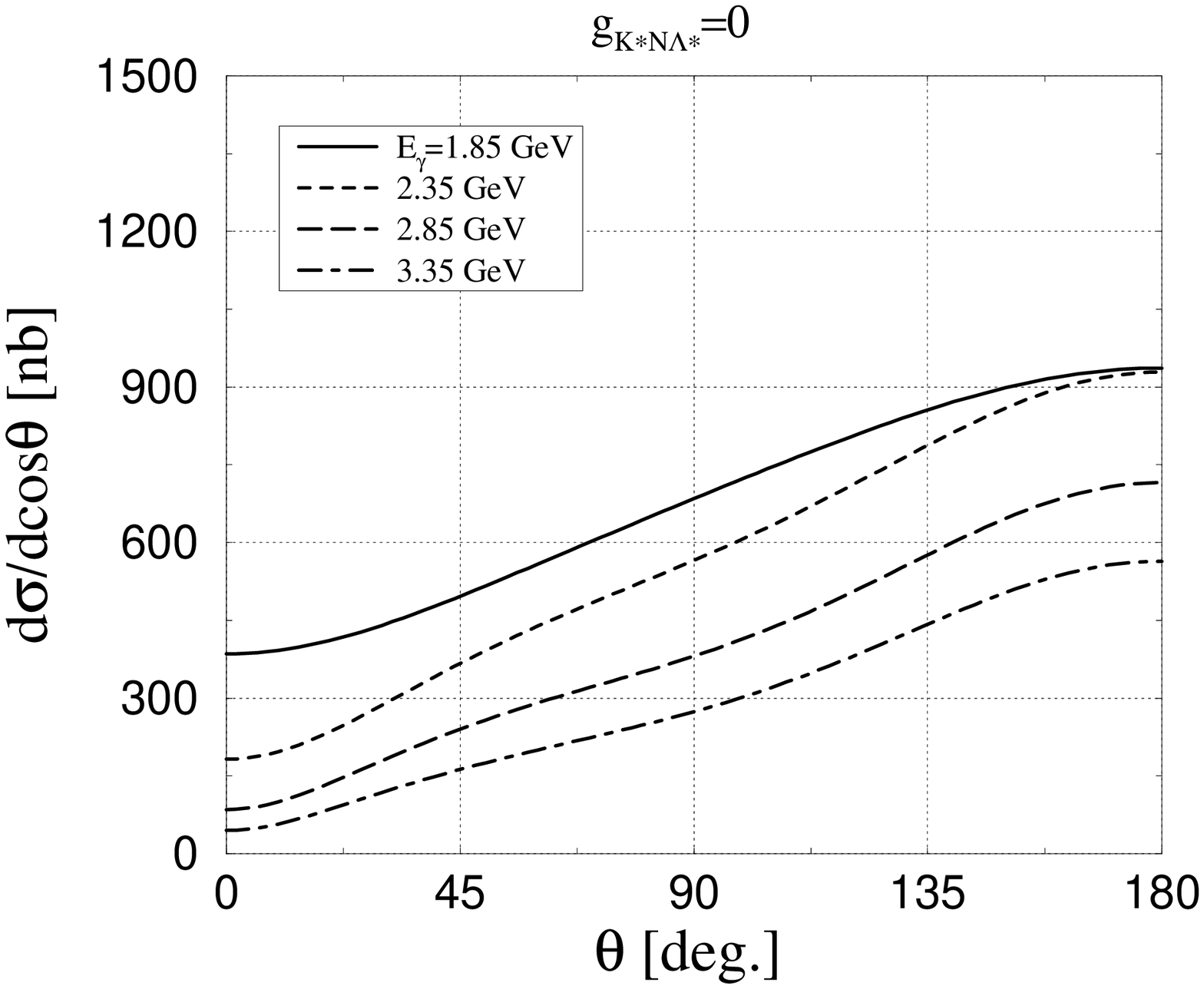}}
\resizebox{5.5cm}{5.5cm}{\includegraphics{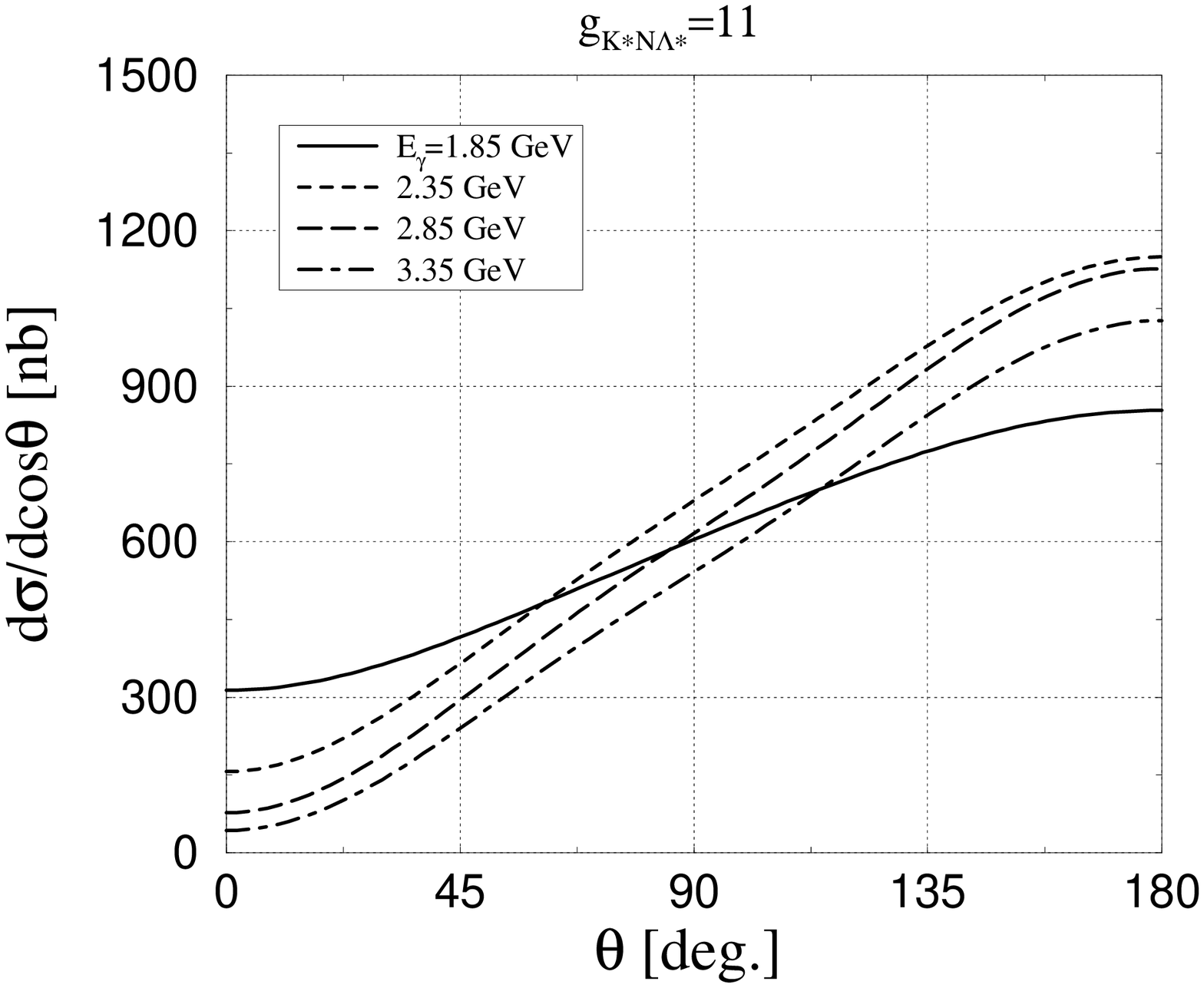}}
\end{tabular}
\caption{The differential cross sections for the proton target with the
form factor $F_2$.  Several photon energies are taken into
account.  We choose  $(\kappa_{\Lambda^*},X) = (0,0)$.}
\label{fig9}
\end{figure}  

In Fig.~\ref{fig9}, we depict the differential cross sections for the proton
target at $E_\gamma=3.8$ GeV with the form factor $F_2$.  Compared to
those with $F_1$ drawn in Fig.~\ref{fig5}, they look very different.  When
the $F_2$ is employed, the backward peak is enhanced as the energy
increases, whereas the $F_1$ does the forward one.  These behaviors arise from
the different angular dependences of the form factors.  Note that the
$F_1$ suppresses the differential cross sections at backward angles, 
while the $F_2$ does not influence the angular distribution.  The
ambiguity arising from the form factors is one of the sources of theoretical
uncertainties in describing hadronic reactions, in particular, at the
higher energy region.  By fitting the results to the experimental
data, we can reduce those uncertainties.

\begin{figure}[tbh]
\begin{tabular}{cc}
\resizebox{8cm}{5.5cm}{\includegraphics{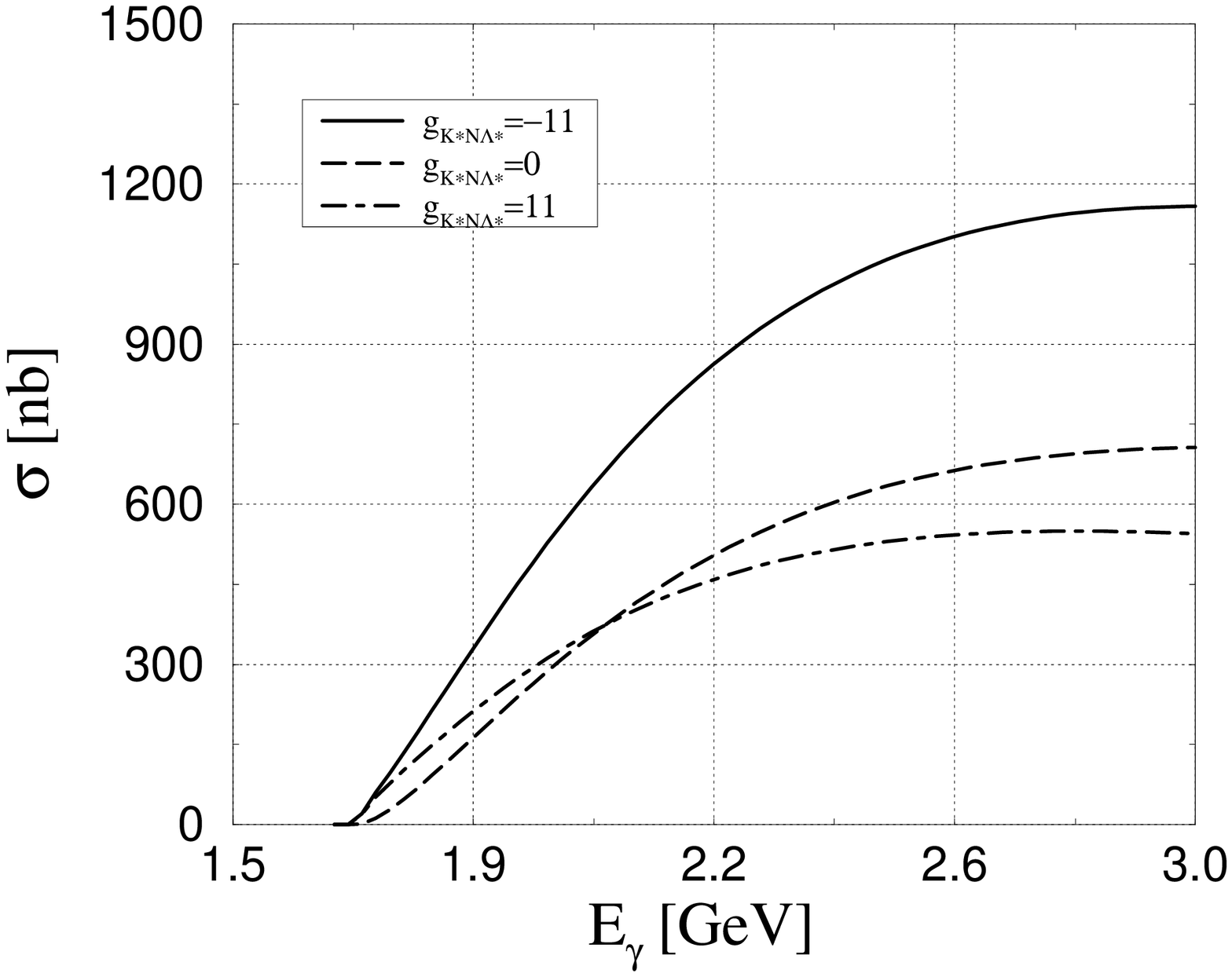}}
\resizebox{8cm}{5.5cm}{\includegraphics{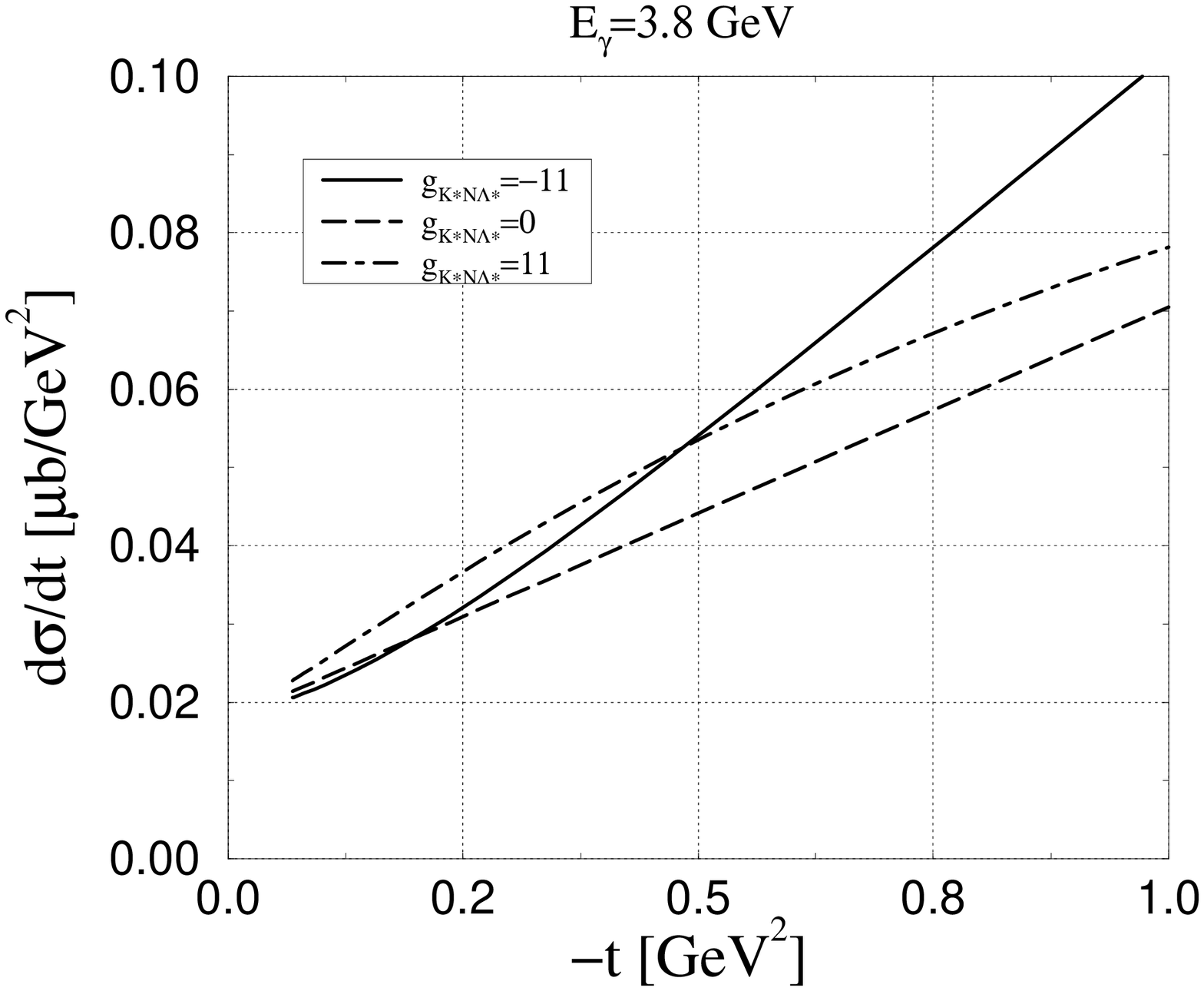}}
\end{tabular}
\caption{In the left panel, the total cross sections are depicted for
  the neutron target with the form factor $F_1$, while in the right
  panel the $t$--dependence is drawn at $E_{\gamma}=3.8$ GeV.  We choose
  $(\kappa_{\Lambda^*},X) = (0,0)$ and three different values of the 
coupling constants, i.e. $g_{K^*N\Lambda^*}=0$ and $\pm 11$.}  
\label{fig10}
\end{figure}

\begin{figure}[tbh]
\begin{tabular}{ccc}
\resizebox{5.5cm}{5.5cm}{\includegraphics{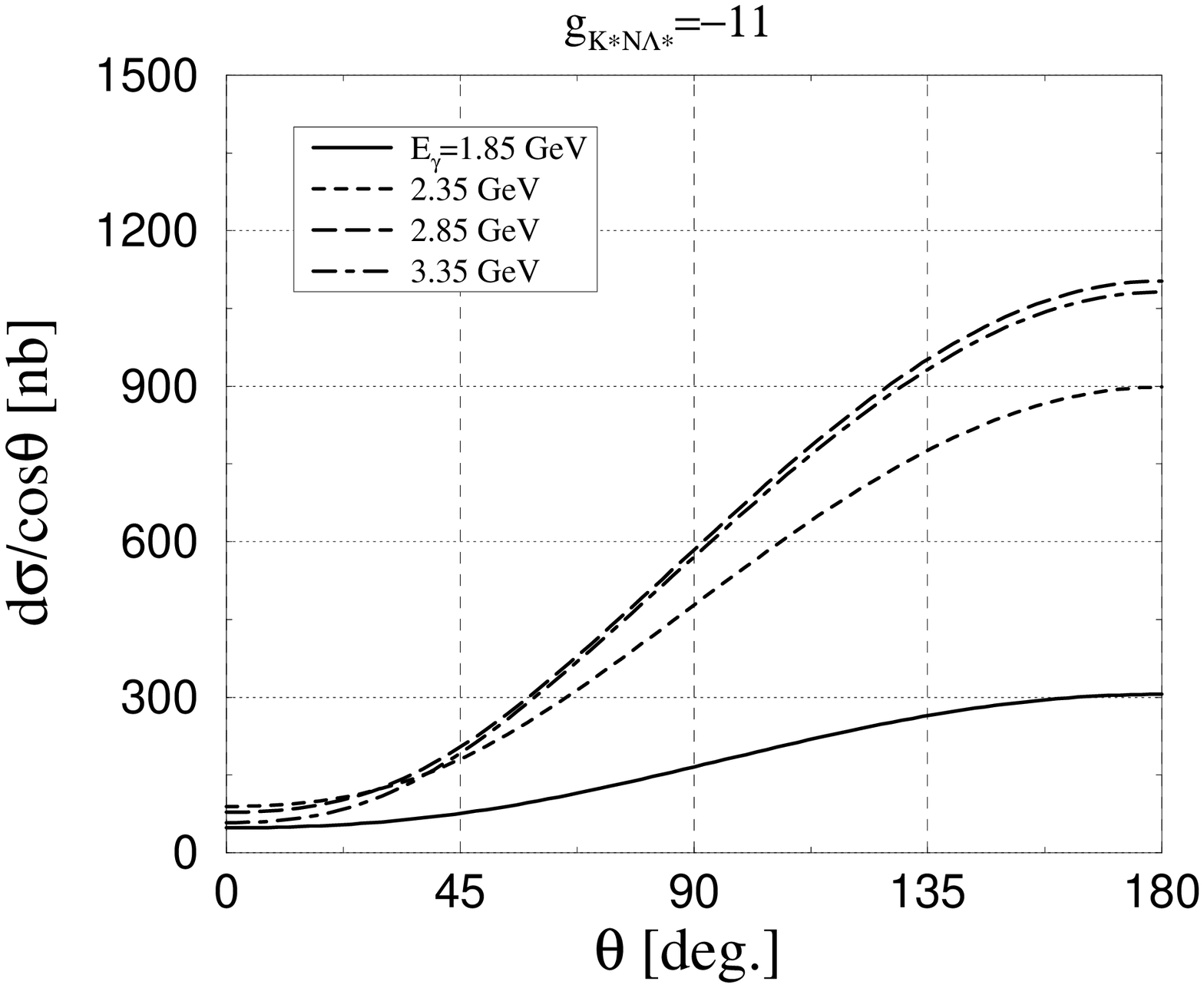}}
\resizebox{5.5cm}{5.5cm}{\includegraphics{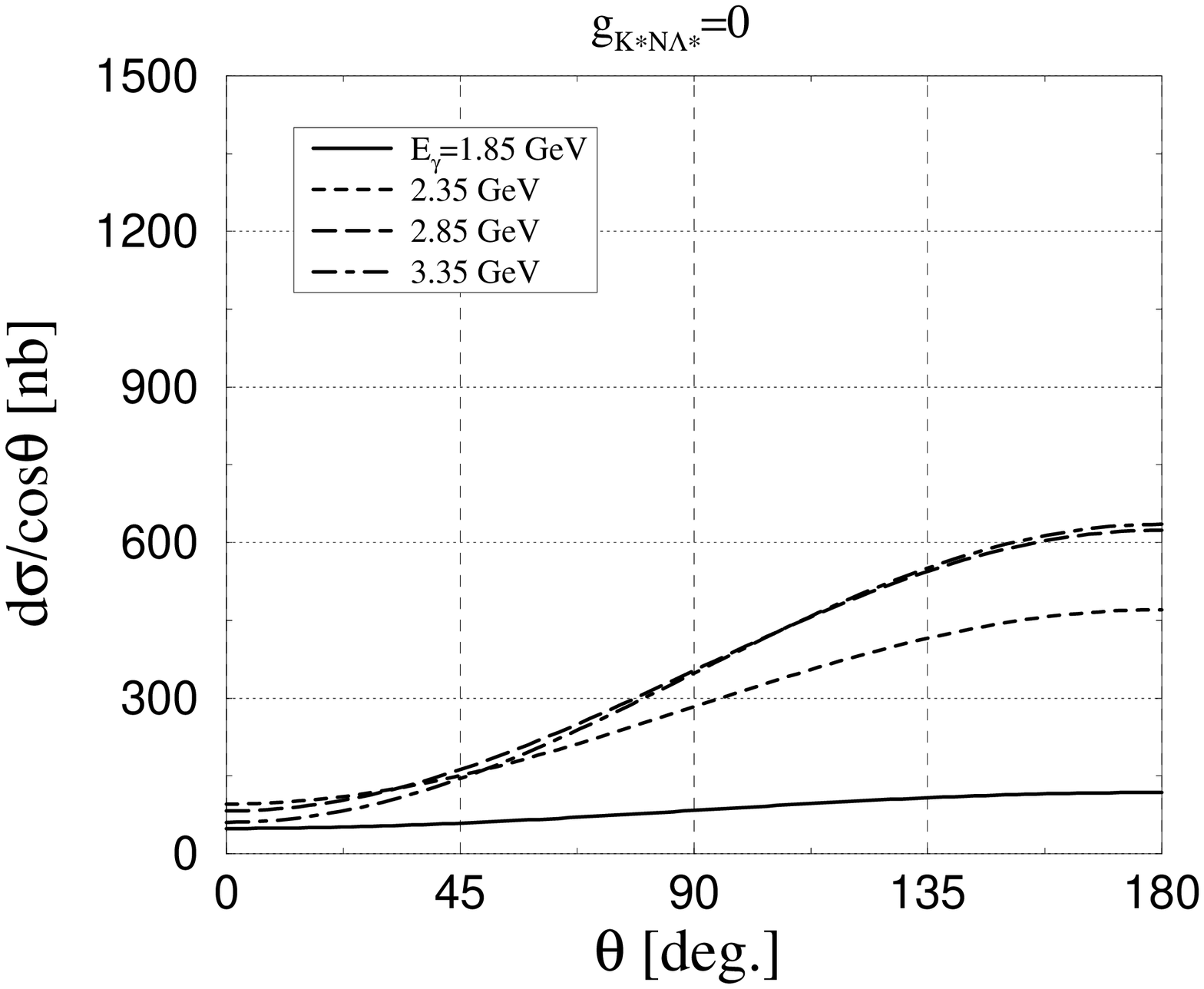}}
\resizebox{5.5cm}{5.5cm}{\includegraphics{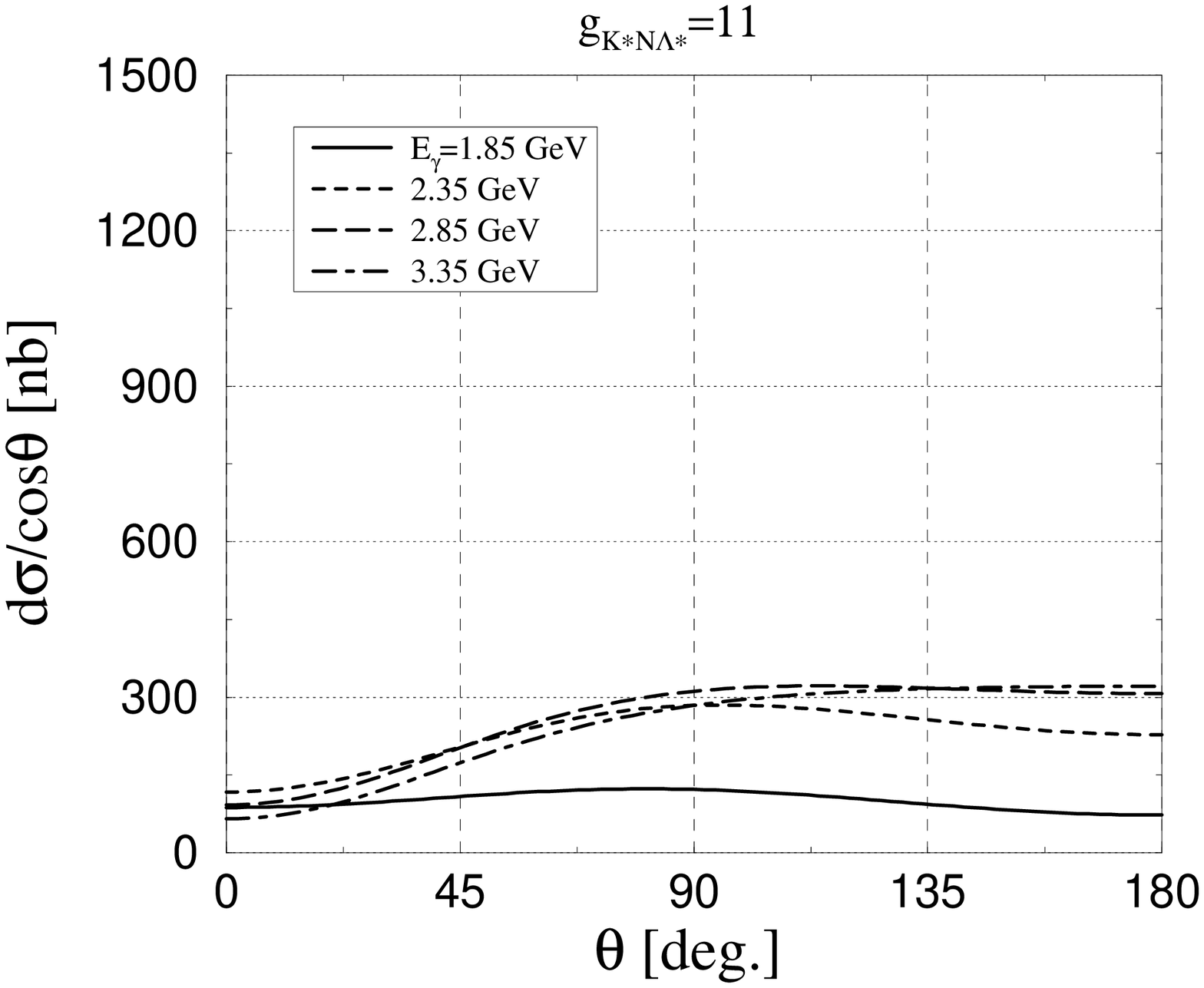}}
\end{tabular}
\caption{The differential cross sections for the neutron target with the
form factor $F_2$.  Several photon energies are taken into
account. We choose $(\kappa_{\Lambda^*},X) = (0,0)$.}
\label{fig11}
\end{figure}

Finally, we discuss the total cross sections for the neutron
target.  The left panel of Fig.~\ref{fig10} draws them and the right
one the $t$--dependence, and Fig.~\ref{fig11} the
angular distribution.  In contrast with those with the form factor
$F_1$, the sign and absolute values of $g_{K^*N\Lambda^*}$ do not
influence much the total cross sections, since the form factor $F_2$
suppresses all channels on the same footing.  Thus, the magnitudes of
the total cross sections are rather similar to those of the proton
target. 
As shown in Fig.~\ref{fig11}, the backward bump is even more enhanced 
as the energy increases in comparison with those for the proton
target.  However, as the energy increases, the size of the bump is
more or less saturated.  
\subsection{Comparison to the $\gamma N\to \bar{K}\Theta^+$ reaction}
Recently, the LEPS collaboration has performed an experiment searching for
the $\Theta^+$ in the two-body process $\gamma d\to
\Lambda^*\Theta^+$.  Since, the statistics for the $\Lambda^*$
photoproduction is much higher than that of the $\Theta^+$, the
reaction can be used to extract information on the production
mechanism of the $\Theta^+$ by comparing it with
with the $\Lambda^*$ photoproduction.  In the previous
studies~\cite{nam2,nam3} and in the present work, we have observed that
the $\gamma n\to K^-\Theta^+$ and $\gamma p\to K^+\Lambda^*$
reactions are less parameter--dependent.  Note that both are charge-exchange
processes.   We first consider the positive--parity $\Theta^+$ with
spin-$1/2^+$.  We  
apply the gauge-invariant from-factor $F_1$ with the cutoff mass
$\Lambda=750$ to both reactions.  The coupling constants for the 
$\Theta^+$ are taken to be $g_{KN\Theta}=1.0$~\cite{Eidelman:2004wy}
and $g_{K^*N\Theta}=+\sqrt{3}g_{KN\Theta}$~\cite{Close:2004tp}.  The
former corresponds to the decay width  $\Gamma_{\Theta\to KN}\sim
1$ MeV.  Then we find that the total cross section for the $\Theta^+$
photoproduction is $2 \sim 3$ $nb$ in average up to $E_{\gamma}=3.0$
GeV.  The angular dependence shows a bump around $50^{\circ}$ at
$E_{\gamma}=1.8$ GeV. The bump moves closer to $\theta=0$ as the
photon energy increases.  Since the angular dependence
of the $\Lambda^*$ photoproduction is enhanced in the forward
direction as the photon energy increases (see Fig.~\ref{fig5}), the
angular dependences for both reactions are qualitatively similar each
other. Considering the angular dependences of the two  
reactions and the present total cross section of the $\Lambda^*$
photoproduction (see Fig.~\ref{fig3}), we find the ratio $R$ of the
total cross sections of these two reactions as follows:
\be
R=\frac{\sigma_{\gamma n\to K^-\Theta^+}}{\sigma_{\gamma p\to
K^+\Lambda^*}}=\frac{1}{300}\sim  \frac{1}{400}.
\label{ratio}
\ee 

As for the negative--parity $\Theta^+$, the values of $R$ will be
decreased approximately by a factor of $10$~\cite{nam2}.  We 
verify that even if we use the form factor $F_2$, the situation does
not change much.  
\section{The role of $K^*$--exchange}
In Ref.~\cite{Barber:1980zv} of the Daresbury experiment, it was
argued that the $\Lambda^*$ photoproduction 
was dominated by vector $K^*$--exchange ($v$--channel) rather than 
pseudoscalar $K$--exchange ($t$--channel) by analyzing the decay
amplitude in the $t$--channel in the helicity basis of the 
$\Lambda^*$. If the helicity of the $\Lambda^*$ is
$S_z=\pm3/2$, the decay of $\Lambda^*\to K^-p$ is explained by
$\sin^2\theta$ in which $\theta$ is the angle between
the two kaons in the helicity basis (see Ref.~\cite{Barrow:2001ds} for
details).  On the other hand, $1/3+\cos^2\theta$ characterizes the
angular dependence of the decay of the $S_z=\pm1/2$ state. Therefore,
taking into account the ratio of these two helicity amplitudes, one
could extract information as to which meson would dominate.  In
Ref.~\cite{Barber:1980zv}, it was shown that the ratio of  
$(S_z=\pm1/2)/(S_z=\pm3/2)$ was nearly zero.  Thus, it was
suggested that the $\Lambda^*$ photoproduction was dominated by the
$v$--channel.  

\begin{figure}[tbh]
\begin{tabular}{ccc}
\resizebox{5.5cm}{5.5cm}{\includegraphics{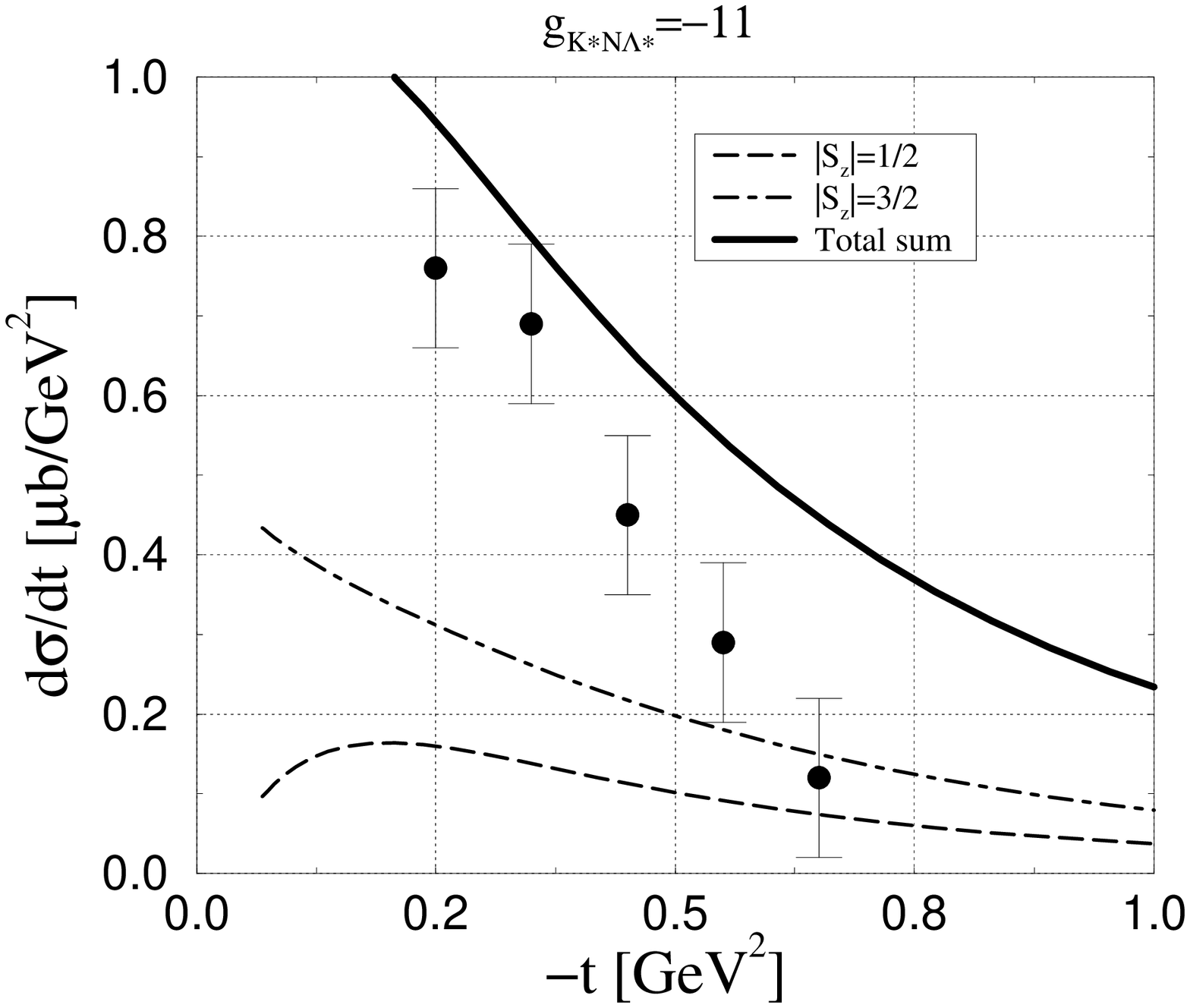}}
\resizebox{5.5cm}{5.5cm}{\includegraphics{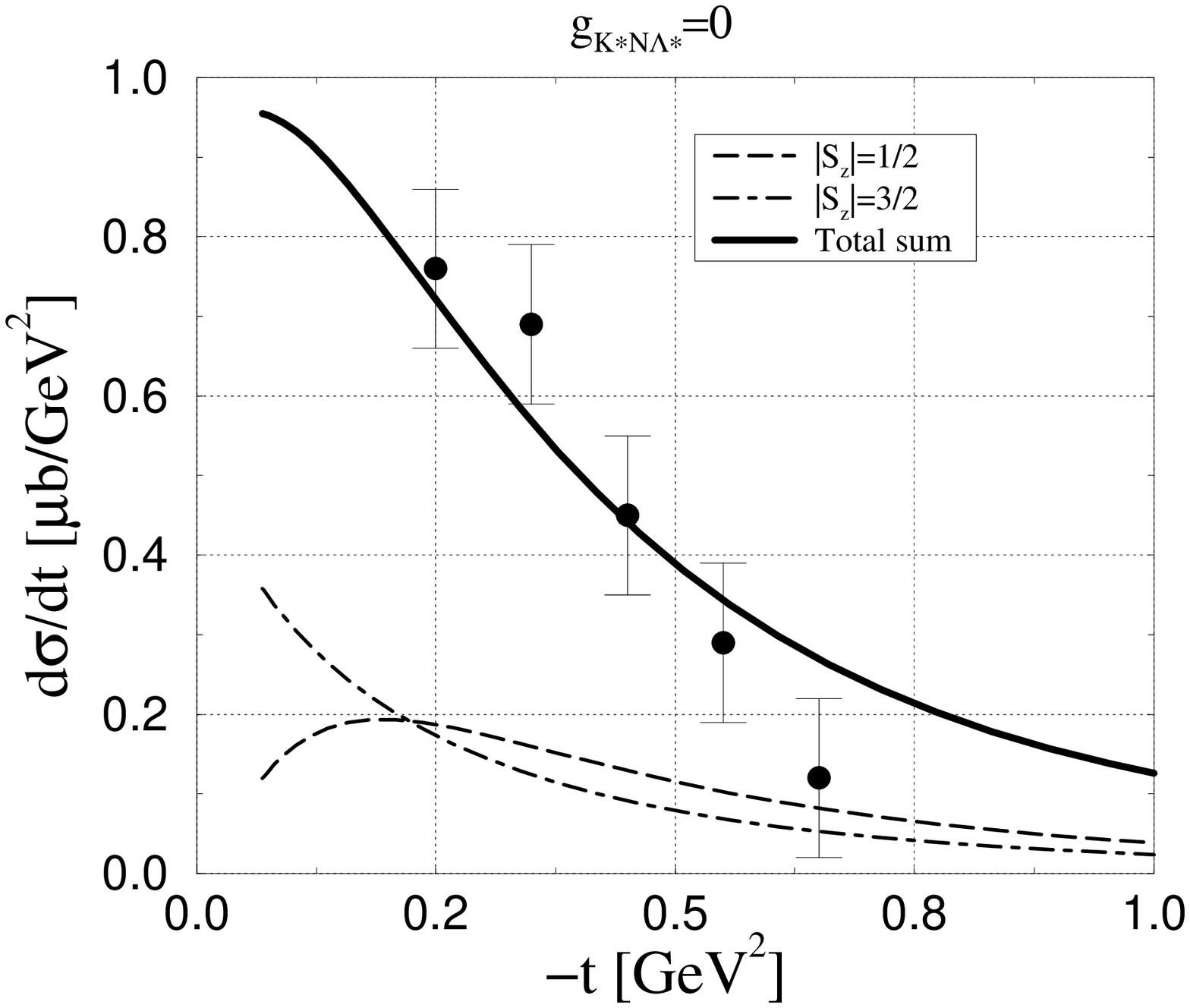}}
\resizebox{5.5cm}{5.5cm}{\includegraphics{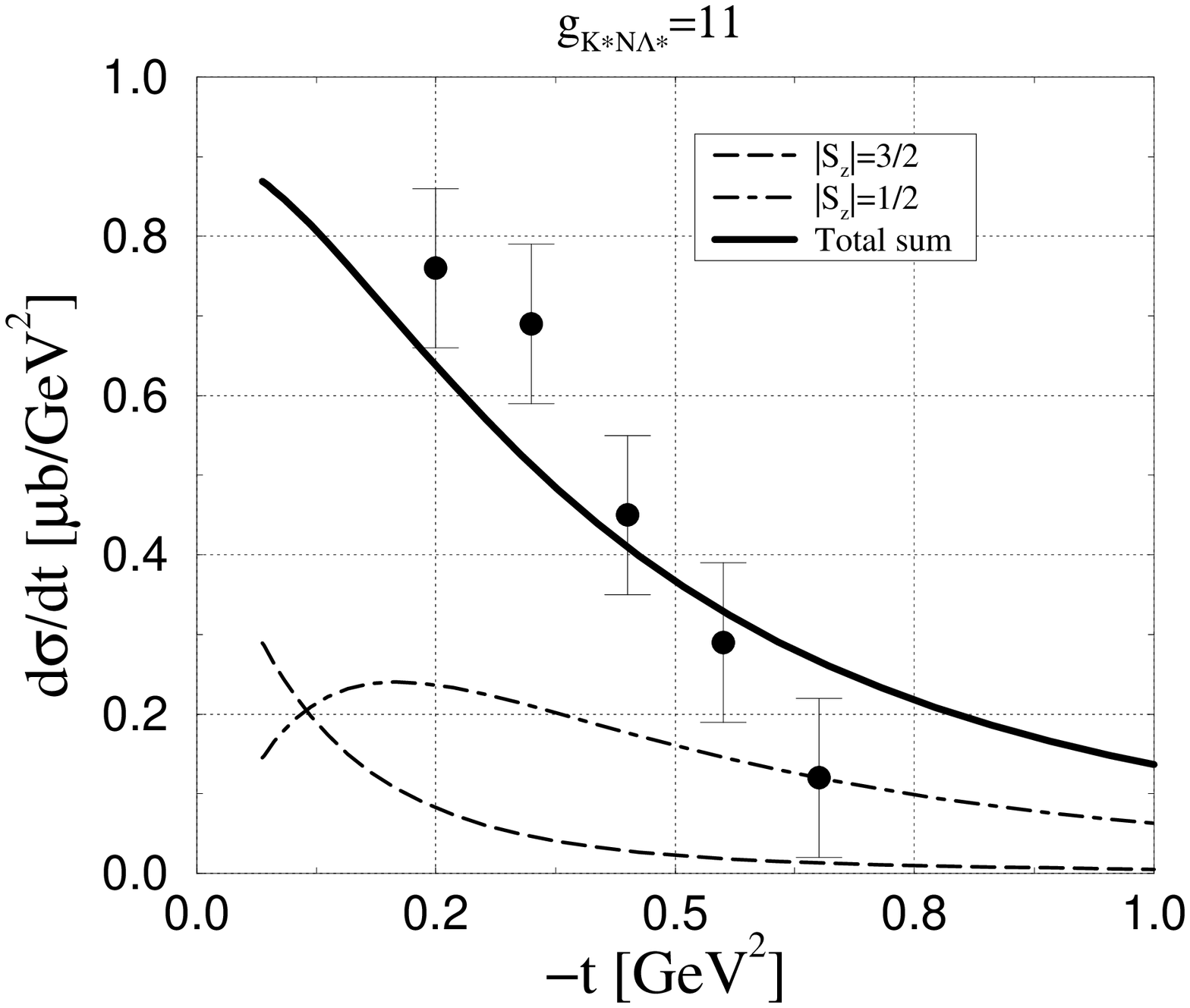}}
\end{tabular}
\caption{The $t$--dependence for each helicity of the $\Lambda^*$ in
the final state.  We change the coupling 
constant $g_{K^*N\Lambda^*}$. We choose  $(\kappa_{\Lambda^*},X) = (0,0)$.}
\label{fig12}
\end{figure}

In Fig.~\ref{fig12}, we plot the $t$--dependence for each helicity using
the form factor $F_1$ with three different values for the coupling
constants $g_{K^*N\Lambda^*}$.  Here, we do not discuss the case of using
the form factor $F_2$, since this form factor fails to reproduce the
experimental data of Ref.~\cite{Barber:1980zv}.  We choose
$E_{\gamma}=3.8$ GeV as done previously.     
In Fig.~\ref{fig12}, we observe that the $S_z=\pm3/2$ contribution is 
dominant especially in the region $-t\lsim0.2\,{\rm GeV}^{-2}$.  There
is also a small contribution from the $S_z=\pm1/2$.  However, we find
that even without the $v$--channel ($g_{K^*N\Lambda^*}=0$)
the $S_z=\pm3/2$ does not become zero.  Therefore, the $S_z=\pm3/2$
contribution comes not only from the $v$--channel but also from the
other channels.  

\begin{figure}[tbh]
\begin{tabular}{ccc}
\resizebox{5.5cm}{5.5cm}{\includegraphics{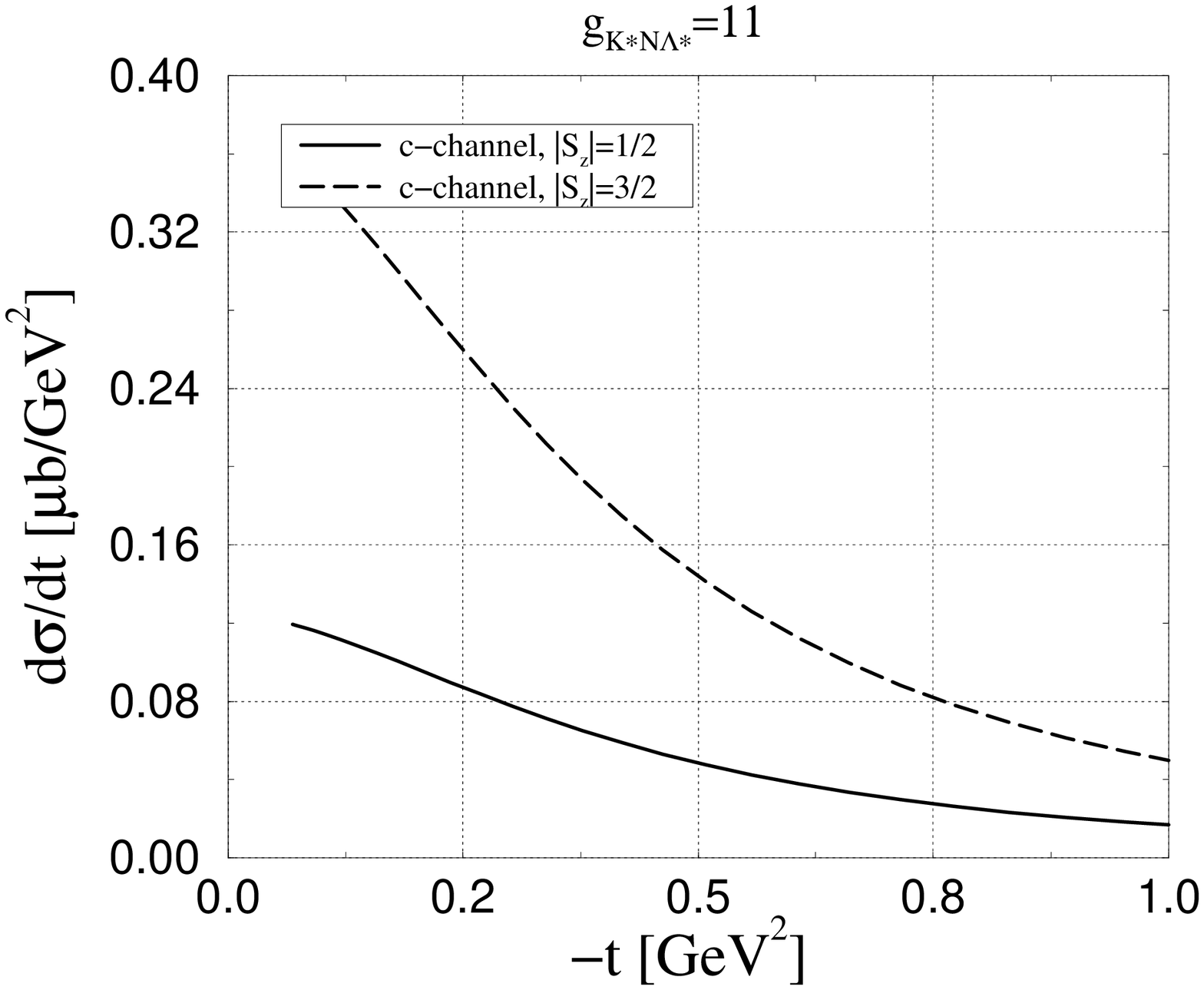}}
\resizebox{5.5cm}{5.5cm}{\includegraphics{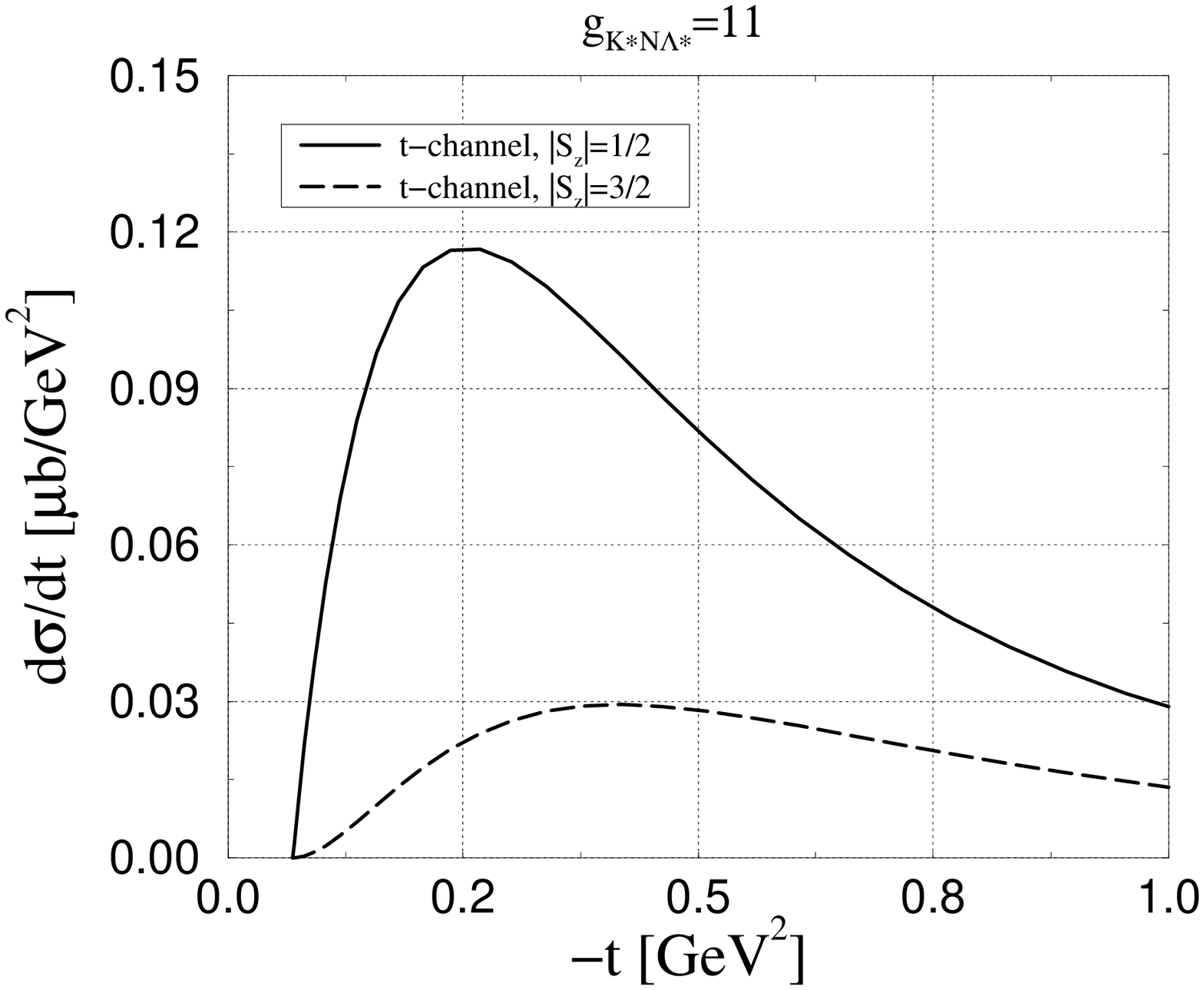}}
\resizebox{5.5cm}{5.5cm}{\includegraphics{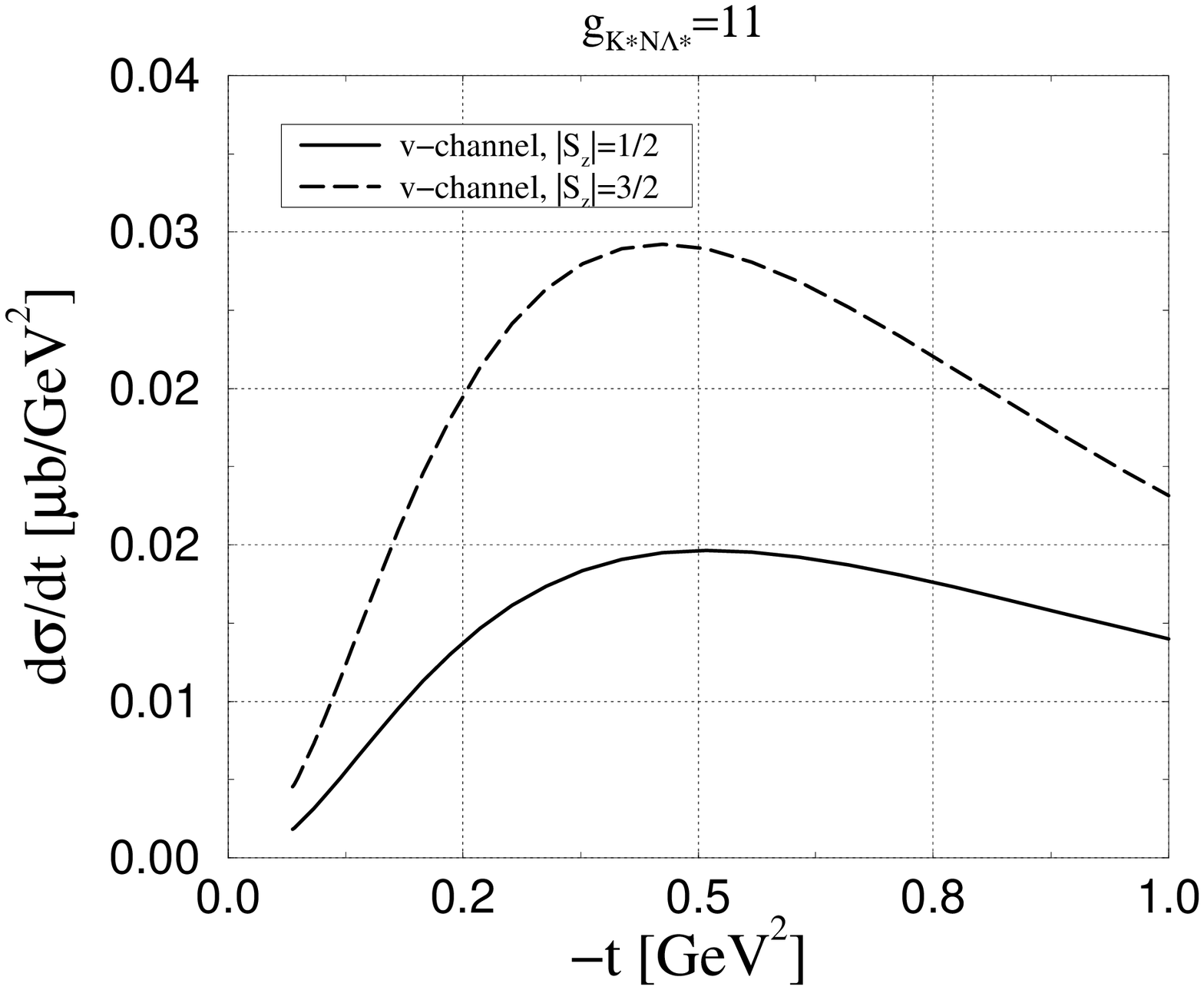}}
\end{tabular}
\caption{The $t$--dependence for the two helicities $S_z=\pm 1/2$ and
  $S_z=\pm 3/2$ for the $c$--,   $t$-- and $v$--channels.}
\label{fig13}
\end{figure}

In order to see this situation more 
carefully, we pick up three important channels, the $c$--,
$t$-- and $v$--channels, and plot the $t$--dependence for each
helicity in Fig.~\ref{fig13}.  One can see that the $S_z=\pm1/2$
contribution is larger than that of the  $S_z=\pm3/2$ for 
pseudoscalar $K$--exchange ($t$--channel), and vice versa for the
$v$--channel.  We also observe that the $c$--channel has sizable
contributions to both $S_z=\pm1/2$ and $S_z=\pm3/2$ amplitudes.  
From these observations, our model calculation using the form factor
$F_1$ indicates that the $S_z=\pm3/2$ contribution is significant 
as shown in Ref~\cite{Barber:1980zv}.  However, most of the 
$S_z=\pm3/2$ contribution comes from the $c$--channel, not from the
$v$--channel as suggested in Ref.~\cite{Barber:1980zv}.  We also find 
that the sizable $S_z=\pm1/2$ contributions are produced from the
$c$-- and $t$--channels.  Therefore, in order to reproduce a nearly
zero value of the ratio of
$(S_z=\pm1/2)/(S_z=\pm3/2)$~\cite{Barber:1980zv}, we need a more   
suppression factor in the $t$--channel, which is the major source of 
the $S_z=\pm1/2$ contribution in the $\Lambda^*$ photoproduction.  
\section{Summary and conclusions}
In the present work, we investigated the $\Lambda^*(1520,3/2^-)$
photoproduction via the $\gamma N\to K\Lambda^*$ reaction.  We
employed the Rarita-Schwinger formalism for describing the spin-3/2
particle for a relativistic description.  Taking into account the
effective Lagrangians for the Born diagrams, we constructed the invariant
amplitudes for the reaction.  The model parameters such as the
anomalous magnetic moment of $\Lambda^*$, $\kappa_{\Lambda^*}$ and the
off-shell parameter $X$ were tested for their sensitivity.  We found that
the parameter dependence turned out to be rather weak in the
low--energy region ($E_{\gamma}\lsim3$ GeV).  Furthermore, the
quark-model calculation indicated that $\kappa_{\Lambda^*}$ was
relatively small and can be ignored. Therefore, we set these two
unknown parameters, $\kappa_{\Lambda^*}$ and $X$, equal to zero for the
numerical calculations.  The coupling constant $g_{K^*N\Lambda^*}$ was 
taken to be $0$ and $\pm11$, since the quark model showed that
$g_{K^*N\Lambda^*}$ was in the same order as $g_{KN\Lambda^*}$.  In
order to check the theoretical ambiguities, we used the gauge-invariant
four-dimensional form factor $F_1$ and the three-dimensional one
$F_2$.  

We performed the numerical calculations for the $\gamma p\to
K^+\Lambda^*$ and $\gamma n\to K^0\Lambda^*$ separately for the
two different types of the form factors.  As for the total cross
sections for the proton target, these two different form factors
gave similar results in magnitude and energy dependence, whereas quite
different behaviors were found for the neutron target.  The total
cross sections for the neutron target using the form factor $F_1$ are
much smaller than those with the $F_2$.  However, since the
$F_2$ failed to reproduce the existing experimental data of the
momentum transfer $t$--dependence for the proton target, it can be
ruled out. The $F_1$ describes it qualitatively well.  We summarize
the whole numerical results briefly in Table.~\ref{table1}.  
\begin{table}[tbh]
\begin{tabular}{|c|c|c|c|c|}
\hline
Form factor&\multicolumn{2}{c|}{$F_1$}&\multicolumn{2}{c|}{$F_2$}\\
\hline
Reactions&$\gamma p\to K^+\Lambda^*$ &$\gamma n\to K^0\Lambda^*$&$\gamma p\to
K^+\Lambda^*$ &$\gamma n\to K^0\Lambda^*$\\
\hline
$\sigma$&$\sim900\,nb$&$\sim30\,nb$&$\sim1200\,nb$&$\sim 700\,nb$\\
${d\sigma}/{d(\cos\theta)}$&Forward peak& Peak at
$\sim45^{\circ}$&Backward peak&Backward peak\\
${d\sigma}/{dt}$&Good&No data&Bad&No data\\
\hline
\end{tabular}
\caption{Summary of the results.}
\label{table1}
\end{table}   

When we compare the present results to those for the $\Theta^+$
photoproduction, the ratio of the total cross sections for these two 
photo-reactions turns out to be $1/300\sim1/400$ for the
positive--parity $\Theta^+$ baryon.  As for the negative parity
$\Theta^+$, it is suppressed by a factor of about
ten~\cite{nam2,nam3}.   We confirm that this ratio is 
less dependent on the model parameters for the charge--exchange
reactions such as $(\gamma n\to K^-\Theta^+,\,\gamma p\to
K^+\Lambda^*)$.  As for the helicity dependence, though the 
contribution of the $S_z=\pm3/2$ was dominant, it was not directly
related to the $K^*$--exchange dominance as suggested in in
Ref.~\cite{Barber:1980zv}.  
 
\section*{Acknowledgments}
We thank T.~Nakano, H.~Toki, A.~Titov, E.~Oset and M.~J.~Vicente Vacas for
fruitful discussions and comments. The work of S.I.N. has been
supported by the scholarship from the Ministry of Education, Culture, 
Science and Technology of Japan. The works of A.H. and S.I.N. are
partially supported by the collaboration program of RCNP, Osaka Univ.,
Japan and IFIC, Valencia Univ., Spain. The work of A.H. is also
supported in part by the Grant for Scientific Research ((C)
No.16540252) from the Education, Culture, Science and Technology of
Japan. The works of H.Ch.K. and S.I.N. are supported by the Korean
Research Foundation (KRF--2003--070--C00015).    
\section*{Appendix}

\subsection{Rarita-Schwinger vector-spinor}
We can write the RS vector-spinors according to their spin states as follows: 
\bee
u^{\mu}(p_{2},\frac{3}{2})&=&e^{\mu}_{+}(p_{2})u(p_{2},\frac{1}{2}),\nn\\
u^{\mu}(p_{2},\frac{1}{2})&=&\sqrt{\frac{2}{3}}e^{\mu}_{0}(p_{2})u(p_{2},\frac{1}{2})
+\sqrt{\frac{1}{3}}e^{\mu}_{+}(p_{2})u(p_{2},-\frac{1}{2}),\nn\\
u^{\mu}(p_{2},-\frac{1}{2})&=&\sqrt{\frac{1}{3}}e^{\mu}_{-}(p_{2})u(p_{2},\frac{1}{2})
+\sqrt{\frac{2}{3}}e^{\mu}_{0}(p_{2})u(p_{2},-\frac{1}{2}),\nn\\
u^{\mu}(p_{2},-\frac{3}{2})&=&e^{\mu}_{-}(p_{2})u(p_{2},-\frac{1}{2}).
\eee
Here, we employ the basis four-vectors, $e^{\mu}_{\lambda}$ which are
written by 
\bee
&&e^{\mu}_{\lambda}(p_{2})=\left(\frac{\hat{e}_{\lambda}\cdot\vec{p}_2}{M_{B}},\,\,\,\,\,\hat{e}_{\lambda}
+\frac{\vec{p}_{2}(\hat{e}_{\lambda}\cdot\vec{p}_{2})}{M_{B}(p^{0}_{2}+M_{B})}\right)\,{\rm
  with}\nn\\
&&\hat{e}_{+}=-\frac{1}{\sqrt{2}}(1,i,0),\,\,\,\,\,\,\hat{e}_{0}=(0,0,1)\,\,\,\,\,\,{\rm and}\,\,\,\,\,\hat{e}_{-}
=\frac{1}{\sqrt{2}}(1,-i,0).
\eee 
\\



\begin{thebibliography}{99}
\bibitem{experiment}
T.~Nakano {\it et al.}  [LEPS Collaboration],
Phys.\ Rev.\ Lett.\  {\bf 91}, 012002 (2003); V.~V.~Barmin {\it et al.}  [DIANA Collaboration],
Phys.\ Atom.\ Nucl.\  {\bf 66}, 1715 (2003)
[Yad.\ Fiz.\  {\bf 66}, 1763 (2003)]; S.~Stepanyan {\it et al.}  [CLAS Collaboration],
Phys.\ Rev.\ Lett.\  {\bf 91}, 252001 (2003); V.~Kubarovsky {\it et al.}  [CLAS Collaboration],
Erratum-ibid.\  {\bf 92}, 049902 (2004)
[Phys.\ Rev.\ Lett.\  {\bf 92}, 032001 (2004)]; J.~Barth {\it et al.}  [SAPHIR Collaboration],
hep-ex/0307083; A.~Airapetian {\it et al.}  [HERMES Collaboration],
Phys.\ Lett.\ B {\bf 585}, 213 (2004).
\bibitem{Nakano:chiral05}
T.~Nakano, talk in the international workshop Chiral 05, RIKEN,
February (2005). 
\bibitem{Boyarski:1970yc}
A.~Boyarski, R.~E.~Diebold, S.~D.~Ecklund, G.~E.~Fischer, Y.~Murata, B.~Richter and M.~Sands,
Phys.\ Lett.\ B {\bf 34}, 547 (1971).
\bibitem{Barber:1980zv}
D.~P.~Barber {\it et al.},
Z.\ Phys.\ C {\bf 7}, 17 (1980).
\bibitem{Barrow:2001ds}
S.~P.~Barrow {\it et al.}  [Clas Collaboration],
Phys.\ Rev.\ C {\bf 64}, 044601 (2001).
\bibitem{Ohta:ji}
K.~Ohta, Phys.\ Rev.\ C {\bf 40}, 1335 (1989).
\bibitem{Haberzettl:1998eq}
H.~Haberzettl, C.~Bennhold, T.~Mart and T.~Feuster, Phys.\ Rev.\ C
{\bf 58}, 40 (1998). 
\bibitem{Davidson:2001qs}
R.~M.~Davidson and R.~Workman, nucl-th/0101066
\bibitem{Rarita:mf}
W.~Rarita and J.~S.~Schwinger,
Phys.\ Rev.\  {\bf 60}, 61 (1941).
\bibitem{Read:ye}
B.~J.~Read,
Nucl.\ Phys.\ B {\bf 52}, 565 (1973).
\bibitem{Johnson:1960vt}
K.~Johnson and E.~C.~G.~Sudarshan,
Annals Phys.\  {\bf 13}, 126 (1961).
\bibitem{Velo:ur}
G.~Velo and D.~Zwanziger,
Phys.\ Rev.\  {\bf 188}, 2218, (1969).
\bibitem{Pascalutsa:1998pw}
V.~Pascalutsa,
Phys.\ Rev.\ D {\bf 58}, 096002 (1998).
\bibitem{Hoehler:gb}
G.~Hoehler, H.~P.~Jakob and R.~Strauss,
Nucl.\ Phys.\ B {\bf 39},  237(1972).
\bibitem{Nath:wp}
L.~M.~Nath, B.~Etemadi and J.~D.~Kimel,
Phys.\ Rev.\ D {\bf 3}, 2153,  (1971).
\bibitem{Hagen:ea}
C.~R.~Hagen,
Phys.\ Rev.\ D {\bf 4}, 2204  (1971).

\bibitem{Machleidt:1987hj}
R.~Machleidt, K.~Holinde and C.~Elster,
Phys.\ Rept.\  {\bf 149}, 1 (1987).
\bibitem{gourdin}
M. Gourdin, Nuovo Cimento 36, 129 (1965); and, 40A, 225 (1965). 
\bibitem{Eidelman:2004wy}
S.~Eidelman {\it et al.}  [Particle Data Group],
Phys.\ Lett.\ B {\bf 592}, 1 (2004).
\bibitem{nam1}
S.~I.~Nam, A.~Hosaka and H.~-Ch.~Kim, talk in the few-body workshop at
RCNP (2004), hep-ph/0502143. 
\bibitem{nam2}
S.~I.~Nam, A.~Hosaka and H.~-Ch.~Kim,
Phys.\ Lett.\ B {\bf 579}, 43 (2004).
\bibitem{nam3}
S.~I.~Nam, A.~Hosaka and H.~-Ch.~Kim,
arXiv:hep-ph/0403009.
\bibitem{Close:2004tp}
F.~E.~Close and J.~J.~Dudek,
Phys.\ Lett.\ B {\bf 586}, 75 (2004).
\end{thebibliography}
\end{document}